\documentclass[prb,twocolumn,showpacs,amsmath,amssymb]{revtex4-1}

\usepackage[dvipdfmx]{graphicx}

\usepackage{bm}

\usepackage{color}
\usepackage{braket}
\usepackage{physics}

\begin{document}

\title{Multifractality and Fock-space localization in many-body localized states:\\
one-particle density matrix perspective}
\author{Takahiro Orito}
\author{Ken-Ichiro Imura}
\affiliation{Graduate School of Advanced Science and Engineering, Hiroshima University, 739-8530, Japan}

\date{\today}

\begin{abstract}
Many-body localization (MBL) is well characterized in Fock space.
To quantify the degree of this Fock space localization,
the multifractal dimension $D_q$ is employed;
it has been claimed that
$D_q$ shows a jump from
the delocalized value $D_q=1$ in the ETH phase (ETH: eigenstate thermalization hypothesis)
to a smaller value $0<D_q<1$ at the ETH-MBL transition,
yet exhibiting a conspicuous discrepancy from the fully localized value $D_q=0$, 
which indicate that multifractality remains
inside the MBL phase.
Here, to better quantify the situation we employ,
instead of the commonly used computational basis,
the one-particle density matrix (OPDM)
and
use its eigenstates (natural orbitals)
as a Fock state basis for representing
many-body eigenstates $|\psi\rangle$ of the system.
Using this basis,
we compute $D_q$ and other indices quantifying the Fock space localization,
such as the local purity $S$, 
which is derived from the occupation spectrum $\{n_\alpha\}$ (eigenvalues of the OPDM).
We highlight the statistical distribution of Hamming distance $x_{\mu\nu}$ occurring
in the pair-wise coefficients $|a_\mu|^2|a_\nu|^2$ 
in $S$, and compare this with a related quantity considered in the literature.
\end{abstract}

\maketitle

\section{Introduction}
A many-body, i.e., interacting system tends to thermalize under
its own dynamics,
\cite{ETH1,ETH2,ETH3,ETH4}
and at weak disorder
realizes
delocalized eigenstates.
However,
in the regime of strong disorder,
the so-called
local integrals of motion (LIOMs)
\cite{Abanin,LIOM2,LIOM3}
are emergent
and hinder thermalization and transport in the system,
leading the system to a many-body localization (MBL) phase.
The existence of such an intriguing phase
was first suggested theoretically,\cite{BAA}
then supported by
experiments mainly in cold-atom systems.
\cite{MBL_cold_exp1,MBL_cold_exp2,MBL_exp1,MBL_exp2,MBL_exp3}
Emergence of the LIOMs in the MBL phase
leads to
various unusual properties of the MBL phase, 
\cite{ImbrieScar_andp}
such as
Poisson level statistics,\cite{Huse1} 
area-law behavior of the entanglement entropy,\cite{Bauer_2013,khemaniPRX}
and its very slow (logarithmic) spreading in time,\cite{Znidarc,log1,log2} 
etc.
While the Anderson localization \cite{PWA58}
for a non-interacting system
occurs in the real space,
MBL can be regarded as localization in the Fock space.
\cite{FSL1,Logan2019,Logan2020}
After intensive study in the last decade
both from theoretical and experimental sides,
the basic understanding on the physics of MBL
has now been established.
\cite{MBL_review1,MBL_review2,MBL_review3,Abanin_RMP}

In the regime of weak disorder 
delocalized eigenstates follow
the eigenstate thermalization hypothesis (ETH)
\cite{ETH1,ETH2,ETH3,ETH4}; 
i.e.,
the eigenstates are also delocalized in the Fock space,
realizing effectively a micro-canonical ensemble;
under such a circumstance
the expectation value of a local observable, e.g., the local magnetization,\cite{Luitz_PRB2016}
takes a well-defined thermodynamic value
in a given energy window between $E$ and $E+\Delta E$.
In the MBL phase, on the other hand,
realized eigenstates involve only a fraction of the available Fock space,
and in the extremely localized limit, 
the eigenstates become a simple product 
of LIOM orbitals, i.e.,
the Anderson localization orbitals dressed by the interaction.
\footnote{
In this limit,
only a certain single coefficient $a_{\mu_j}(=1)\neq 0$ 
in Eq. (\ref{beta_0})
is finite, and all others vanish
($a_{\mu}=0$ for $\mu\neq\mu_j$),
where
$\mu=\{n_1,n_2,\cdots\}$.
}
In such a MBL phase,
the system is no longer in equilibrium, and the expectation value of local observables
fluctuate.\cite{Luitz_PRB2016}
To quantify such different situations
in the ETH and MBL phases,
one considers
the inverse participation ratio (IPR)
in the Fock space [defined in Eq. (\ref{IPR_F})],
or a related quantity, the multifractal dimension $D_q$
[defined in Eq. (\ref{mfd})]. 
Note that
$D_q=1$ for a fully delocalized state, while
$D_q=0$ for a fully localized state, 
and
the intermediate situation:
$0<D_q<1$ is called multifractal.
In non-interacting higher dimensional systems
delocalization-localization occurs at a single point,
and only at this point the system becomes multifractal ($0<D_q<1$).
Here, in 1D interacting systems the situation is rather different;
preceding works
\cite{mirlin_MFD,Tarzia_MFD,Luitz_MFD,Laflo_M, Pollmann_HD} 
have suggested
that $D_q=1$ in the ETH phase, while after the ETH-MBL transition
$D_q$ remains 
multifractal ($0<D_q<1$), 
reflecting
the many-body nature of the system;
i.e.,
at the ETH-MBL transition
$D_q$ does not show a complete transition to the ideal value
$D_q=0$ corresponding to true localization
as far as the disorder strength is finite.
Still, $D_q$ shows a partial discontinuity at the ETH-MBL transition, 
and a similar discontinuity is also expected in the entanglement entropy.
\cite{DeT,Pollmann_HD}
In Refs. \onlinecite{Laflo_C,Pollmann_HD} 
the meaning of the finiteness of $D_q$ in the MBL phase has been analyzed,
and its relation to
the nature of ETH-MBL phase transition is discussed; 
the latter is claimed to be KT-like.\cite{KT1}
In the MBL phase
$D_q$ also strongly fluctuates, and said to be non self-averaging.\cite{santos}
In the avalanche scenario, 
proposed in Refs. \onlinecite{ava1,ava2,ava3}
the multifractalty in the MBL phase
may be given the following natural interpretation:
in a generic situation in the MBL phase
LIOMs are formed, but some ``spins'' are still active
in the pseudospin picture;
i.e., LIOMs are not precisely good quantum numbers.
It is natural to presume that
under such circumstances
a many-body eigenstate 
is only partially localized in the Fock space
(IPR$\neq 1$, $D_q\neq 0$).
When disorder is no longer strong enough,
the density of active spins reaches a certain threshold value,
at which an avalanche of active spins occurs,
destroying (melting) completely the frozen LIOMs,
resulting in the ETH situation:
IPR$\simeq 0$, $D_q\simeq 1$.

Here, in the remainder of the paper we focus on this
intriguing partial localization in the Fock space
in the generic MBL phase.
To what extent 
a many-body eigenstate $|\psi\rangle$ in Eq. (\ref{psi_comp})
is localized
in the Fock space
depends on the basis 
one employs for representing
$|\psi\rangle$.
In numerics,
one {\it a priori} employs the computational basis (\ref{basis_comp}), in which
the coefficients
$a_{\{n_j\}}$ 
in Eq. (\ref{psi_comp})
show a rather broad distribution;
i.e., $|\psi\rangle$ is not much localized
in the corresponding Fock space
even in the MBL phase
and even in the theoretical LIOM limit.
To quantify
the degree of Fock-space localization in the MBL phase
more properly,
it is ideal to employ the basis of LIOM orbitals,
but this is not straightforward, since in a generic MBL situation 
LIOMs are coupled to a thermal bath;
not commuting with the total Hamiltonian, 
they are no longer in the strict sense integrals of the motion.\cite{ava2}
Under such circumstances,
instead of seeking for constructing LIOMs,
it may be more realistic to
employ the eigenstates of the
one-particle density matrix (OPDM).\cite{Bardarson_PRL,Bardarson_andp}
Under an assumption in the deep MBL phase (see Sec. II-B)
the eigenvectors of OPDM, called natural orbitals,
are shown to 
coincide with the LIOMs.
In a more generic situation in the MBL phase
they are assumed to be still good approximations of the LIOMs.
The OPDM approach has been employed in the study of MBL
in various models
\cite{opdm_DMRG,OPDM_attractive,OPDM-quasi,OPDM_zero,orito2020effects,OPDM_bose}
and in the study of
out-of-equilibrium phenomena.
\cite{OPDM_quench,OPDM_quench2}

In this work,
we have computed
$D_q$ and other indices quantifying the Fock space localization
in the OPDM and other bases, and have compared the results.
With the use of OPDM basis, 
mimicking the LIOM basis,
one can remove, or at least minimize effects of the finiteness
of Fock-space localization length,
which manifests, e.g., in the finiteness of $D_q$ in the MBL phase.
We expect that this will result in
a better description of the ETH-MBL transition/crossover regime.
Our analyses in the OPDM basis
show
that the finiteness of $D_q$ in the computational basis
reported in the literature
is indeed due to the finiteness of the Fock-space localization length.
The eigenenergies $\rho_\alpha$ of the OPDM (occupation spectrum)
shows a characteristic gapped distribution in the MBL phase,
reminiscent of 
a renormalized Fermi distribution 
in Fermi liquids.
\cite{Bardarson_PRL}
In the idealized LIOM case,
this becomes a simple step function as in Fermi gas,
indicating that the corresponding many-body state $|\psi\rangle$ can be
expressed by a single Slater determinant.
Thus, 
the degree of Fock-space localization
is encoded in 
how close the occupation spectrum is to a simple step function.
Or,
one can numerate
this resemblance to a step function
by a single index,
called the local purity.\cite{viola}

The remainder of the paper is organized as follows.
In Sec. II we highlight various aspects of the OPDM 
approach to many-body localization
with a particular emphasis on
the behavior of
occupation spectrum and Fock-space IPR.
and,
in Sec. III
we introduce the quantity called local purity, an index quantifying 
the nature of occupation spectrum.
We compare the behavior of
the local purity with that of the Fock-space IPR
from the viewpoint of
the distribution of Hamming distance $x_{\mu\nu}$
in the pairwise coefficients $|a_\mu|^2|a_\nu|^2$.
Sec. IV is devoted to Concluding Remarks.
Some details are postponed to three sections in the Appendices.

\section{The OPDM approach to MBL}

To fix the notation let us first introduce our model: 
\begin{eqnarray}
H=\sum_{j=1}^{L}\Big[-t(c_{j+1}^{\dagger}c_{j}+c_j^{\dagger}c_{j+1})
+W_{j}\left(\hat{n}_j-\frac{1}{2}\right)
\nonumber\\
+V\left(\hat{n}_j-\frac{1}{2}\right)\left(\hat{n}_{j+1}-\frac{1}{2}\right)\Big],
\label{ham}
\end{eqnarray}
where $j$ represents a site in real space, and
$L$ is the size of the system.
$c_j^{\dagger}$ ($c_j$) creates (annihirates) an electron at site $j$,
and $\hat{n}_j=c_j^{\dagger}c_j$ counts 
the local electron density at site $j$. 
In the first two terms of Eq. (\ref{ham}),
$t$ represents the strength of hopping between the
nearest-neighbor sites, and
in the third term
the strength $W_j$ of the on-site impurity potential is a random variable
at each site $j$ 
and each
obeys the uniform distribution of magnitude $W$;
$W_j \in [-W, W]$.
In the second line
$V$ represents the strength of nearest-neighbor interaction.
The system prescribed by Eq. (\ref{ham}) represents one of the
paradigmatic models for describing the many-body localization phenomenon.
\cite{Abanin_RMP}
The on-site potential term of strength $W$
tends to localize the electronic wave functions, while
the hopping and the interaction terms, each parametrized, respectively, by $t$ and $V$,
tend to delocalize them.
The competition of these three different types of contributions,
each represented by the parameters, $t$, $W$ and $V$,
determine the localization/delocalization feature of the system (cf. {e.g.}, the phase diagram of Ref. \onlinecite{PD_Laflo}.
Note that
in Eq. (\ref{ham}) we presume a periodic boundary condition so that
$c_{j+L}\equiv c_j$.

\begin{figure*}
\includegraphics[width=170mm]{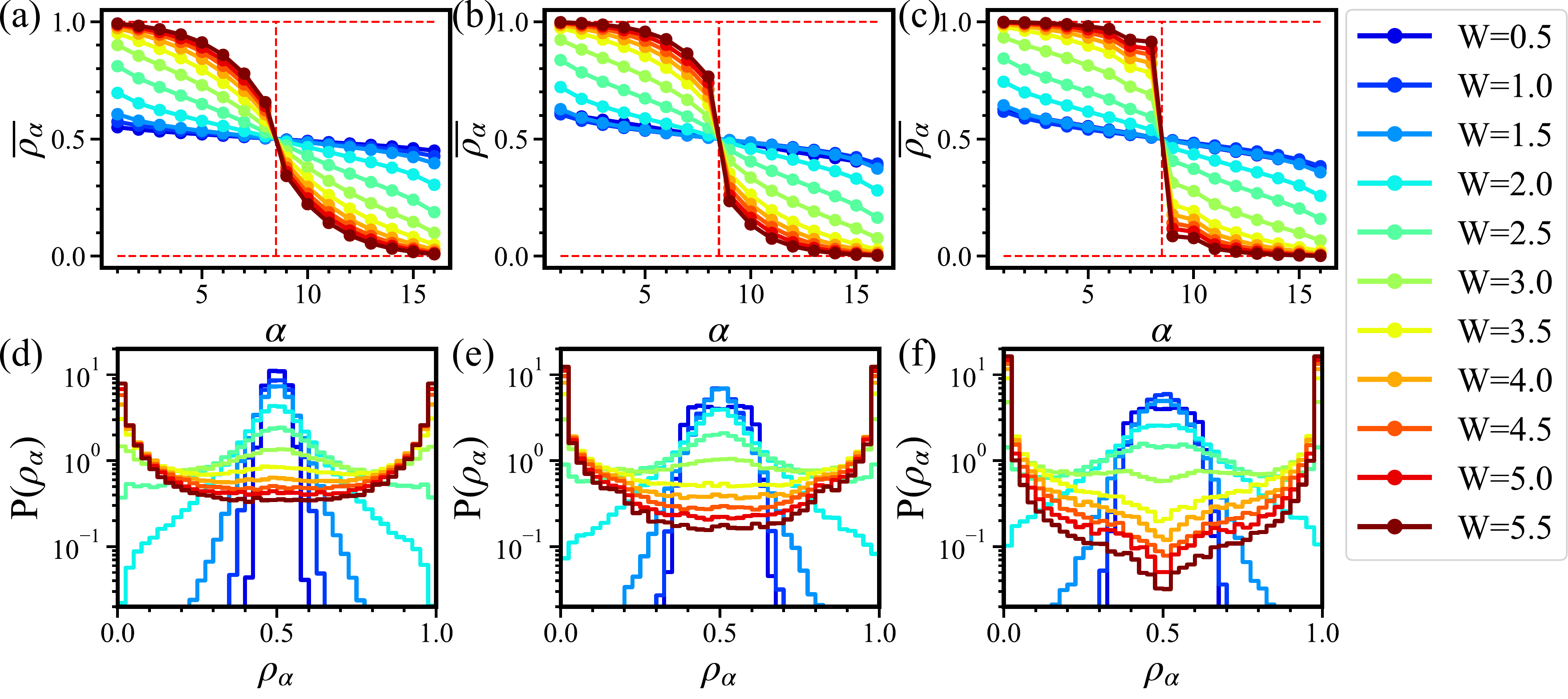}
\caption{The occupation spectrum $\{\overline{\rho_\alpha}\}$
[panels (a)-(c)]
and the corresponding probability distribution $P(\rho_\alpha)$
[panels (d)-(f)]
in different many-body bases
[(a), (d): the computational,
(b), (e): the AL orbital, and
(c), (f): the OPDM
bases].}
\label{occ_spec}
\end{figure*}


A generic many-body eigenstate $|\psi\rangle$ 
of the Hamiltonian
such as Eq. (\ref{ham}), satisfying
$H |\psi\rangle = E |\psi\rangle$,
takes the following form:
\begin{eqnarray}
|\psi\rangle&=&\sum_{\{n_j\}} a_{\{n_j\}} |\{n_j\}\rangle,
\label{psi_comp}
\end{eqnarray}
i.e., a superposition of
\begin{equation}
N=\frac{L!} {N_e!(L-N_e)!}
\label{dim_H}
\end{equation}
different electron configurations:
\begin{eqnarray}
|\{n_j\}\rangle
&\equiv&
|n_1 n_2 \cdots n_L\rangle
\nonumber \\
&=&
(c_{L}^\dagger)^{n_L}
\cdots 
(c_{2}^\dagger)^{n_2}
(c_{1}^\dagger)^{n_1}
|0\rangle,
\label{basis_comp}
\end{eqnarray}
with a suitable weight $a_{\{n_j\}}$.
The notation $\{n_j\}$
specifies a Fock representation:
\begin{equation}
\{n_j\}=(n_1, n_2,\cdots ,n_L),
\end{equation}
where
$n_j=0,1$ (fermionic statistics), and 
$c_j^\dagger$ creates an electron
on a site $j$.
$\sum_{j=1}^L n_j=N_e$ represents the number of electrons.
In numerics,
the many-body basis (\ref{basis_comp})
is usually employed; therefore dubbed as computational basis.
In the following
we focus on the typical case of half-filling: $N_e=L/2$.
\footnote{
At half-filling: $\nu=N_e/L=1/2$,
the dimension $N$ of the many-body Hilbert space becomes
maximal for a given $L$.
For $L=16$
the summation $\sum_{\{n_j\}}$ in Eq. (\ref{psi_comp})
should be taken
over $N=L!/(L/2)!^2=12870\simeq 10^4$ different realizations
of the basis states (\ref{basis_comp}),
and this number increases rapidly with increasing the system size $L$
({\it e.g.}, $N=48620\simeq 5\times 10^4$ for $L=18$).
In the present-day computor performance
a simple diagonalization of the Hamiltonian
such as the one given in Eq. (\ref{H_mn})
can be done up to the size of $L=18$ 
possibly with the help of shift-invert method
within a reasonable duration of order $\Delta t \sim 10^3$ secs.
In this work
much of computation time has been spent
for the calculation of the coefficients $a_{[\alpha]}$
in different many-body bases.
As a result,
the maximal system size considered in this work
has been limited to $L=16$.
}
In numerical simulations
we also set the parameters at the following
typical values: $t=1/2$ and $V=1$.

\subsection{The OPDM and its eigenvalues (the occupation spectrum)}

For a {\it given} 
many-body eigenstate $|\psi\rangle$ 
we introduce a one-particle density matrix (OPDM) $\rho$
whose $(i,j)$-element $\rho_{ij}$ is defined as
\begin{equation}
\rho_{ij}=\langle \psi | c_i^{\dagger}c_j |\psi\rangle,
\label{rho_ij}
\end{equation}
where
$i, j=1, 2, \cdots, L$
represent a site in real space.
We then diagonalize the $L\times L$ matrix $\rho$ (i.e., the OPDM) so that
\begin{equation}
\rho u^{(\alpha)} = \rho_\alpha u^{(\alpha)}.
\label{rho_eigen}
\end{equation}
The set of eigenvalues $\rho_\alpha$
is called the occupation spectrum,
while
the corresponding eigenstates
are called natural orbitals $u^{(\alpha)}$
(for reasons that will become clear below).
\cite{Bardarson_PRL,Bardarson_andp}


Creating an electron in the $\alpha$th natural orbital,
\begin{equation}
u^{(\alpha)}
= (u^{(\alpha)}_1, u^{(\alpha)}_2,\cdots ,u^{(\alpha)}_L)^t
\label{u_alpha}
\end{equation}
can be represented by a creation operator,
\begin{equation}
c_\alpha^{\dagger} = \sum_{j=1}^L  u^{(\alpha)*}_ j c_j^\dagger.
\label{c_alpha}
\end{equation}
The occupation of the $\alpha$th natural orbital 
in the many-body eigenstate $|\psi\rangle$
is specified by the quantity:
 \begin{equation}
\langle \psi | c_\alpha^{\dagger} c_\alpha
|\psi\rangle
=\sum_{ij} u^{(\alpha)*}_i u^{(\alpha)}_j \langle \psi | c_i^{\dagger}c_j |\psi\rangle,
\end{equation}
but
recalling the definitions of the OPDM $\rho$ and of the natural orbitals $u_\alpha$
[Eq. (\ref{rho_ij}) and Eq. (\ref{rho_eigen})],
one immediately finds
that this is identical to $\rho_\alpha$ given in Eq. (\ref{rho_eigen}).
Thus, the set of eigenvalues,
\begin{equation}
\{\rho_\alpha\}=(\rho_{\alpha_1}, n_{\alpha_2},\cdots ,\rho_{\alpha_L})
\end{equation}
of the OPDM $\rho$
specifies how $L$ natural orbitals $u^{(\alpha)}$
are occupied in the state $|\psi\rangle$.

The occupation spectrum $\{\rho_\alpha\}$ computed in the OPDM (natural orbital)
basis is shown in
Fig. \ref{occ_spec} (c).
The set of OPDM eigenvalues
$\{\rho_\alpha\}$
is obtained by numerically diagonalizing the matrix $\rho$ 
for a state $|\psi\rangle$,
then we have labelled them in the descending 
order of $\rho_\alpha$ such that
\begin{equation}
\rho_{\alpha_1}>\rho_{\alpha_2}>\cdots>\rho_{\alpha_L}.
\label{n_order}
\end{equation}
We repeat this procedure for different eigenstates $|\psi\rangle$
in the middle of the spectrum, 
\footnote{to avoid the effect of mobility edges that may appear 
near the top and bottom of the band}
then for different disorder configurations.
Each component $\rho_\alpha$ in the set $\{\rho_\alpha\}$
is then averaged over different samples:
$\rho_\alpha\rightarrow\overline{\rho_\alpha}$, 
and
the sample-averaged occupation spectrum:
\begin{equation}
[\overline{\rho_\alpha}]=\left(
\overline{\rho_{\alpha_1}},\overline{\rho_{\alpha_2}},\cdots,\overline{\rho_{\alpha_L}}
\right)
\end{equation}
is found, where
$\overline{\cdots}$ represents sample averaging. 
In Fig. \ref{occ_spec} (c)
we have repeated this calculation for different disorder strength $W$,
and 
at each value of $W$
we have plotted
the sample-averaged occupation spectrum
$[\overline{\rho_\alpha}]$
as a function of $\alpha$.
At each value of $W\ge 2.5$,
we have averaged in total
over 
$5\times 10^3$ samples.

In the deep
MBL regime: $W\gg 3.5$,
$\rho_\alpha$ shows a sharp jump $\Delta \rho_\alpha$ 
from $\alpha=L/2$ to $\alpha=L/2+1$;
the entire 
shape of the spectrum is close to the form of a step function;
i.e.,
\begin{eqnarray}
\rho_\alpha &=& \theta(\alpha-L/2)
\nonumber \\
&=&
\left\{
\begin{array}{ll}
1  & ({\rm for}\ \alpha> L/2) \\
0  & ({\rm for}\ \alpha\le L/2)
\end{array}
\right..
\label{step}
\end{eqnarray}
As $W$ decreases, the magnitude of the jump $\Delta \rho_\alpha$ diminishes,
and the spectrum $\rho_\alpha$
tends to become a smooth function
that varies only in a small range of values around  $\rho_\alpha=0.5$
in the ETH regime: $W\ll3.5$.

How close the shape of the occupation spectrum is to a step function (\ref{step})
is a measure of 
how close the given many-body eigenstate
$|\psi\rangle$ 
is to a simple product state; 
i.e.,
to what extent the state is Fock-space localized
in the basis chosen.
If the state
$|\psi\rangle$ is expressed in some basis
as a simple product state: 
\begin{equation}
|\psi\rangle=
\gamma_{\beta_{L/2}}^\dagger
\cdots 
\gamma_{\beta_2}^\dagger
\gamma_{\beta_1}^\dagger 
|0\rangle
\equiv |[\beta_0]\rangle_{\rm LIOM},
\label{beta_0}
\end{equation}
where
$\gamma_\beta^\dagger$
creates an electron in the $\beta$th orbital in this basis,
and if one measures
the occupation spectrum $\rho_\beta$
in the same basis,
then
the occupation
\begin{equation}
\rho_\beta=
\langle \psi | \gamma_\beta^{\dagger} \gamma_\beta
|\psi\rangle
\end{equation}
becomes
a simple step function as Eq. (\ref{step}),
since in this case
$\{\rho_\beta\}$ reduces to the simple Fock representation: 
\begin{eqnarray}
\{n_\beta\} 
&=&(n_{\beta_1}, \cdots, n_{\beta_{L/2}}, n_{\beta_{L/2+1}}, \cdots, n_{\beta_L})
\nonumber \\
&=&(1, \cdots, 1, 0, \cdots, 0)
\label{n_beta}
\end{eqnarray}
of the state $|[\beta_0]\rangle_{\rm LIOM}$;
i.e.,
$n_\beta=1$ if $\beta$ is occupied,
while $n_\beta=0$ othehrwise.
The last line holds if
the orbitals $\beta$ are arranged 
in the descending order of $n_\beta$.

\subsection{Relation to LIOM, comparison with other bases}

In the deep MBL regime in which
the local integrals of motion (LIOMs) become good quantum numbers,
the many-body eigenstate
$|\psi\rangle$ 
can 
be
expressed as a single Slater determinant
as in Eq. (\ref{beta_0})
in terms of the LIOM creation operators:\cite{Bardarson_andp}
\begin{equation}
\gamma_\alpha^{\dagger} = \sum_{i}  A^{(\alpha)*}_i c_i^\dagger +
\sum_{ijk}  B^{(\alpha)*}_{ijk} c_i^\dagger c_j^\dagger c_k + \cdots,
\label{LIOM}
\end{equation}
where
$A^{(\alpha)}_j$ represents the principal part of the LIOM wave function,
while
$B^{(\alpha)*}_{ijk}$
represents a correction associate with a particle-hole excitation.
$\cdots$ represents higher-order corrections that stem from
higher-order terms in the perturbative expansion of LIOM.
Here, we consider
the extremely localized limit, and hypothesize
that only the first term of Eq. (\ref{LIOM})
is relevant,
and neglect the terms of order higher than two.\cite{Pollmann_HD}
Then, it is natural to assume
that the amplitudes $A^{(\alpha)}_j$ are orthonormal:
\begin{equation}
\sum_i A^{(\alpha)*}_i A^{(\beta)}_i=\delta_{\alpha\beta},\ \
\sum_\alpha A^{(\alpha)}_i A^{(\alpha)*}_j=\delta_{ij},
\label{LIOM_ortho}
\end{equation}
since they are simply LIOM wave functions.
Using Eq. (\ref{LIOM_ortho}),
one can invert
Eq. (\ref{LIOM}) as
\begin{equation}
c_i^\dagger \simeq \sum_{\alpha}  A^{(\alpha)}_i  \gamma_\alpha^{\dagger}.
\label{c_gamma}
\end{equation}
Then,
the OPDM matrix $\rho$ [Eq. (\ref{rho_ij})]
becomes
\begin{eqnarray}
\rho_{ij}&=&
\langle \psi | c_i^\dagger c_j |\psi\rangle
\nonumber \\
&\simeq&\sum_{\alpha\beta}
A^{(\alpha)}_i A^{(\beta)*}_j
\langle \psi | \gamma_\alpha^{\dagger} \gamma_\beta
|\psi\rangle
\nonumber \\
&=&\sum_{\alpha}
n_\alpha
A^{(\alpha)}_i A^{(\alpha)*}_j
=\sum_{\alpha: {\rm occupied}}
A^{(\alpha)}_i A^{(\alpha)*}_j.
\label{rho_LIOM}
\end{eqnarray}
In the intermediate 
step,
we have used
\begin{equation}
\langle \psi | \gamma_\alpha^{\dagger} \gamma_\beta
|\psi\rangle
=n_\alpha \delta_{\alpha\beta},
\end{equation}
where $n_\alpha$'s are as given in Eq. (\ref{n_beta}).
Note that
Eq. (\ref{rho_LIOM}) is nothing but the spectral decomposition of
the OPDM matrix $\rho$ such that
\begin{equation}
\rho_{ij}=
\sum_{\alpha}
\rho_\alpha
u^{(\alpha)}_i u^{(\alpha)*}_j,
\end{equation}
in the case of $\rho_\alpha=n_\alpha$.
This signifies that
the LIOM orbitals $A^{(\alpha)}_j$
are identical to natural orbitals $u^{(\alpha)}_j$ in this limit,
and
the corresponding occupation spectrum $\{\rho_\alpha\}$
reduces to a simple occupation $\{n_\alpha\}$ [i.e., the Fock representation as given in Eq. (\ref{n_beta})]
of LIOM orbitals;
the latter becomes a simple step function as given in Eq. (\ref{step}).

In a generic situation in the MBL phase,
the LIOM creation operator (\ref{LIOM})
will be still valid,
but higher order terms therein may play some role.
In this case
the natural orbitals are no longer identical to LIOM orbitals
but still close to them,
and the occupation spectrum $\{\rho_\alpha\}$ is no longer
an ideal step function 
but still shows a jump $\Delta \rho_\alpha <1$ from $\alpha=L/2$ to $\alpha=L/2+1$.
Such features can indeed be seen in Fig. \ref{occ_spec} (c)
in the deep MBL regime: $W\gg 3.5$.

Panels (a) and (b) of Fig. \ref{occ_spec}
show the occupation spectrum $\{\rho_\alpha\}$
calculated
(a) in the computational, and
(b) in the AL orbital bases,
for comparison.
In cases (a) and (b),
\begin{eqnarray}
\rho_j
&=&\langle \psi | c_j^\dagger c_j
|\psi\rangle=\rho_{jj},
\nonumber \\
\rho_\alpha^{(AL)}
&=&
\langle \psi | c_{AL}^{(\alpha)\dagger} c_{AL}^{(\alpha)}
|\psi\rangle
\nonumber \\
&=&
\sum_{ij}  \psi_{AL}^{(\alpha)*}(i) \psi_{AL}^{(\alpha)}(j) \rho_{ij} 
\label{rho_ab}
\end{eqnarray}
have been calculated, respectively,
and then sample-averaged,
where
$\rho_{ij}$ is the OPDM matrix, and
\begin{equation}
c_{AL}^{(\alpha)\dagger} = \sum_{j=1}^L  \psi_{AL}^{(\alpha)*}(j)c_j^\dagger,
\label{c_AL}
\end{equation}
creates 
an electron in the $\alpha$th AL orbital $\psi_{AL}^{(\alpha)}$.

In the ETH regime ($W$: small $\ll3.5$)
the occupation spectrum $\rho_\alpha$ becomes
a smooth function
that varies only in a small range of values around $\rho_\alpha=0.5$
in all the three bases;
i.e., 
the local observable $\overline{\rho_\alpha}$ exhibits a well-defined 
thermodynamical value 
at a given energy (for a given $|\psi\rangle$),
realizing a situation
consistent with the hypothesis of ETH.
In the MBL regime ($W$: large $\gg 3.5$),
on the other hand,
the values of $\overline{\rho_\alpha}$ differ on the two sides of the jump at $\alpha=L/2$,
breaking the hypothesis of ETH.
This also implies localization in the Fock space.

In the deep MBL regime: $W\gg 3.5$,
the occupation spectrum $\rho_\alpha$ is closest to a step function
in (c) the OPDM (natural orbital) basis,
indicating that
in this basis
the many-body eigenstate is 
predominantly described by a single Slater determinant.
This also
implies that the corresponding natural orbitals are close to
those of LIOMs.
In cases of (a) and (b),
the occupation spectrum $\rho_\alpha$ deviates significantly 
from a simple step function;
the AL orbitals [case of panel (b)] 
and those of the computational basis; i.e., $\psi_j (x)=\delta(x-j)$,
are not particularly close to
LIOM orbitals.

\subsection{Probability distribution of $\rho_\alpha$}
To see why
the occupation spectrum $\rho_\alpha$ deviates significantly,
especially, in cases of (a) and (b),
from a simple step function,
even in the MBL phase,
we then consider
the (statistical) distribution of $\rho_\alpha$
in each and different realizations;
here
we cease to order $\rho_\alpha$'s in each realization
[as in Eq. (\ref{n_order})]
and focus on the occurrence of the
quantal expectation value 
(\ref{rho_ab}) in cases (a) and (b)
and the eigenvalue $\rho_\alpha$ of the OPDM 
[Eq. (\ref{rho_eigen})]
for $\alpha=1,2,\cdots,L$
and in different realizations
and its distribution
in the range $[0,1]$.
We have counted the number of occurrences $\rho_\alpha$
in the bins $[\rho_\alpha, \rho_\alpha+\Delta \rho_\alpha]$, 
and establish a histogram of $\rho_\alpha$;
then after normalization one finds the distribution $P(\rho_\alpha)$.

In the ETH limit
the distribution $P(\rho_\alpha)$ is expected to be 
a narrow gaussian function centered at $\rho_\alpha=1/2$,
while
in the deep MBL phase
$\rho_\alpha$ takes either 0 or 1;
as a result
$P(\rho_\alpha)$ is expected to become a bimodal function
sharply peaked at 0 and 1.
In Fig. \ref{occ_spec}
we show such distribution $P(\rho_\alpha)$,
(d) in the computational,
(e) in the Anderson localization orbital, and
(f) in the OPDM (natural orbital)
bases.
The width of the bins is chosen as $\Delta \rho_\alpha=1/40$.
At weak $W$,
in the ETH regime: $W\ll3.5$,
the distribution $P(\rho_\alpha)$ shows a peak centered at $\rho_\alpha=1/2$
in all three different bases,
while the peak broadens as $W$ increases.
In the MBL regime: $W\gg 3.5$,
$P(\rho_\alpha)$ becomes a bimodal function peaked at 0 and 1,
but with a relatively long tail,
in cases of (d) and (e),
extended toward the center of the distribution $\rho_\alpha=1/2$.

The reason why
$\rho_\alpha$ takes such values away from the extreme values 0 and 1
is that the LIOM wave function has a finite localization length 
even in the deep MBL phase.
In the LIOM limit
the many-body eigenstate $|\psi\rangle$ is expressed 
by a simple product state 
as in Eq. (\ref{beta_0}),
in which
$\gamma_\alpha^\dagger$
represents a LIOM creation operator as in Eq. (\ref{LIOM}).
Then,
the occupation $\langle n_j \rangle$
in the computational basis can be expressed
as a superposition of LIOM orbitals as
\begin{equation}
\rho_j=
\langle \psi | c_j^\dagger c_j |\psi\rangle \simeq
\sum_{\alpha: {\rm occupied}} |A^{(\alpha)}_j|^2,
\label{n_comp}
\end{equation}
where, for simplicity, we have kept
only the first term of Eq. (\ref{LIOM}),
and have neglected the higher order terms.
In this case
the creation operator $c_j^\dagger$ in the computational basis
can be written 
as in Eq. (\ref{c_gamma}).
The LIOM orbital $A^{(\alpha)}_j$ has a finite spread in real space;
i.e.,
a finite localization length $\xi_\alpha$ such that
\cite{Laflo_C}
\begin{equation}
A^{(\alpha)}_j \simeq 
\sqrt{\tanh \frac{1} {2\xi_{\alpha}}}
\exp \left(-\frac{|j-j_\alpha|}{2\xi_\alpha}\right),
\label{LIOM_exp}
\end{equation}
which indicates that
Eq. (\ref{n_comp})
represents
a superposition of 
exponentially decaying amplitudes 
centered at localization centers $j_\alpha$.
At a generic site $j$,
contributions of
the tails from different localization centers $j_\alpha$ superpose
and give a finite amplitude,
where
$\alpha\in$ occupied states.
The distribution
$P(\langle n_j \rangle)$ in this case
is expected to have a relatively long tail away from the extreme values
$\langle n_j \rangle =0,1$.

In the AL orbital basis,
one measures, instead of $c_j^\dagger$,
$c_{AL}^{(\alpha)\dagger}$
given in Eq. (\ref{c_AL}).
Here, in the LIOM limit,
using Eq. (\ref{c_gamma}),
one rewrites Eq. (\ref{c_AL}) as
\begin{eqnarray}
c_{AL}^{(\alpha)\dagger} &=& \sum_{j=1}^L  \psi_{AL}^{(\alpha)*}(j)c_j^\dagger
\nonumber \\
&=& \sum_{j,\beta} \psi_{AL}^{(\alpha)*}(j) A^{(\beta)}_j  \gamma_\beta^{\dagger}
=\sum_{\beta} 
\tilde{A}^{(\beta)}_\alpha
\gamma_\beta^{\dagger},
\label{c_AL2}
\end{eqnarray}
where
we have introduced the amplitude 
\begin{equation}
\tilde{A}^{(\beta)}_\alpha=\sum_{j} \psi_{AL}^{(\alpha)*}(j) A^{(\beta)}_j.
\label{A_tilde}
\end{equation}
Clearly,
$\tilde{A}^{(\beta)}_\alpha$ represents the overlap of the
$\alpha$th AL orbital and
$\beta$th LIOM orbital.
In the non-interacting limit,
the orbitals coincide so that 
Eq. (\ref{A_tilde}) reduces to an orthogonality relation:
$\tilde{A}^{(\beta)}_\alpha=\delta_{\alpha\beta}$.
In case of $V\neq 0$, $\tilde{A}^{(\beta)}_\alpha$ is no longer $\delta_{\alpha\beta}$,
but is still close.
In the reversed relation:
\begin{equation}
\gamma_\beta^{\dagger}=
\sum_{\alpha}
\tilde{A}^{(\beta)*}_\alpha
c_{AL}^{(\alpha)\dagger},
\end{equation}
$\tilde{A}^{(\beta)}_\alpha$ can be also regarded as
the amplitude of $\beta$th LIOM wave function
in the AL orbital basis $\alpha$.
The LIOM wave function $A^{(\alpha)}_j$ has a finite spread
in the real space basis $j$ as in Eq. (\ref{LIOM_exp}),
while
in the AL orbital basis
$\tilde{A}^{(\beta)}_\alpha$ is closer to a $\delta$-function $\delta_{\alpha\beta}$;
at least it will be more localized than in the real space basis.
Using Eq. (\ref{c_AL2}),
one can express
the occupation $\rho_\alpha^{(AL)}$
of the $\alpha$th AL orbital as
\begin{equation}
\rho_\alpha^{(AL)}=
\langle \psi | c_{AL}^{(\alpha)\dagger}c_{AL}^{(\alpha)}
|\psi\rangle\simeq
\sum_{\beta: {\rm occupied}} |\tilde{A}^{(\beta)}_\alpha|^2,
\label{n_AL}
\end{equation}
i.e., in the form of
a superposition of localized orbitals as in Eq. (\ref{n_comp}).
The difference is that here
each contribution is more localized 
in the space of AL orbitals;
therefore,
the distribution
$P(\rho_\alpha^{(AL)})$
is expected to have a larger weight in the vicinity of two extreme values
$\rho_\alpha^{(AL)}=0,1$.

In the case of OPDM basis,
if one considers the same LIOM limit
under the hypothesis of neglecting the higher order terms
of Eq. (\ref{LIOM}),
then
the natural orbitals $u^{(\alpha)}_j$
are shown to be identical to
LIOM wave functions $A^{(\alpha)}_j$.
As compared to  the case of AL orbital basis,
$\psi_{AL}^{(\alpha)*}(j)$ should be replaced with
$u^{(\alpha)*}_j$
in the expression for $\tilde{A}^{(\beta)}_\alpha$ in Eq. (\ref{A_tilde}).
Then, 
$u^{(\alpha)}_j=A^{(\alpha)}_j$ signifies
\begin{equation}
\tilde{A}^{(\beta)}_\alpha=\sum_{j} u^{(\alpha)*}_j A^{(\beta)}_j
=\delta_{\alpha\beta},
\label{delta}
\end{equation}
i.e.,
the LIOM wave function $\tilde{A}^{(\beta)}_\alpha$
is ultimately localized in the OPDM basis.
Conferring to Eq. (\ref{n_AL}),
this implies that
the distribution
$P(\rho_\alpha)$ in the OPDM basis becomes
a ultimately sharp bimodal function peaked at $\rho_\alpha=0,1$.
Of course, in reality
the higher order terms of Eq. (\ref{LIOM}) play some role,
so that
Eq. (\ref{delta}) 
does not literally hold.
As a result,
$P(\rho_\alpha)$ still has some weights (though much suppressed)
away from the extreme values $\rho_\alpha=0,1$.

In the MBL phase,
$P(\rho_\alpha)$ is a U-shaped function in cases of (a) and (b),
showing a broad minimum around $\rho_\alpha=1/2$,
while
in case (c),
as $W$ increases,
a dip evolves at $\rho_\alpha=1/2$,
deforming the global shape
to V-shaped.
This explains
why in Fig. \ref{occ_spec}
the occupation spectrum $\{\rho_\alpha\}$ 
becomes a sharp step function
in case (c)
in the deep MBL regime,
while
the step is washed out in cases (a), (b).
The U-shaped $P(\rho_\alpha)$ 
in cases (a), (b)
has a non-negligible amplitude at and around $\rho_\alpha=1/2$
enough to wash out the step in the occupation spectrum $\{\rho_\alpha\}$
at $\alpha=L/2$ to $\alpha=L/2+1$,
while 
such contributions are exponentially supressed in case (c);
note the semi-log scale in the plots in Fig. \ref{occ_spec} (d-f).

\begin{figure}
\includegraphics[width=85mm]{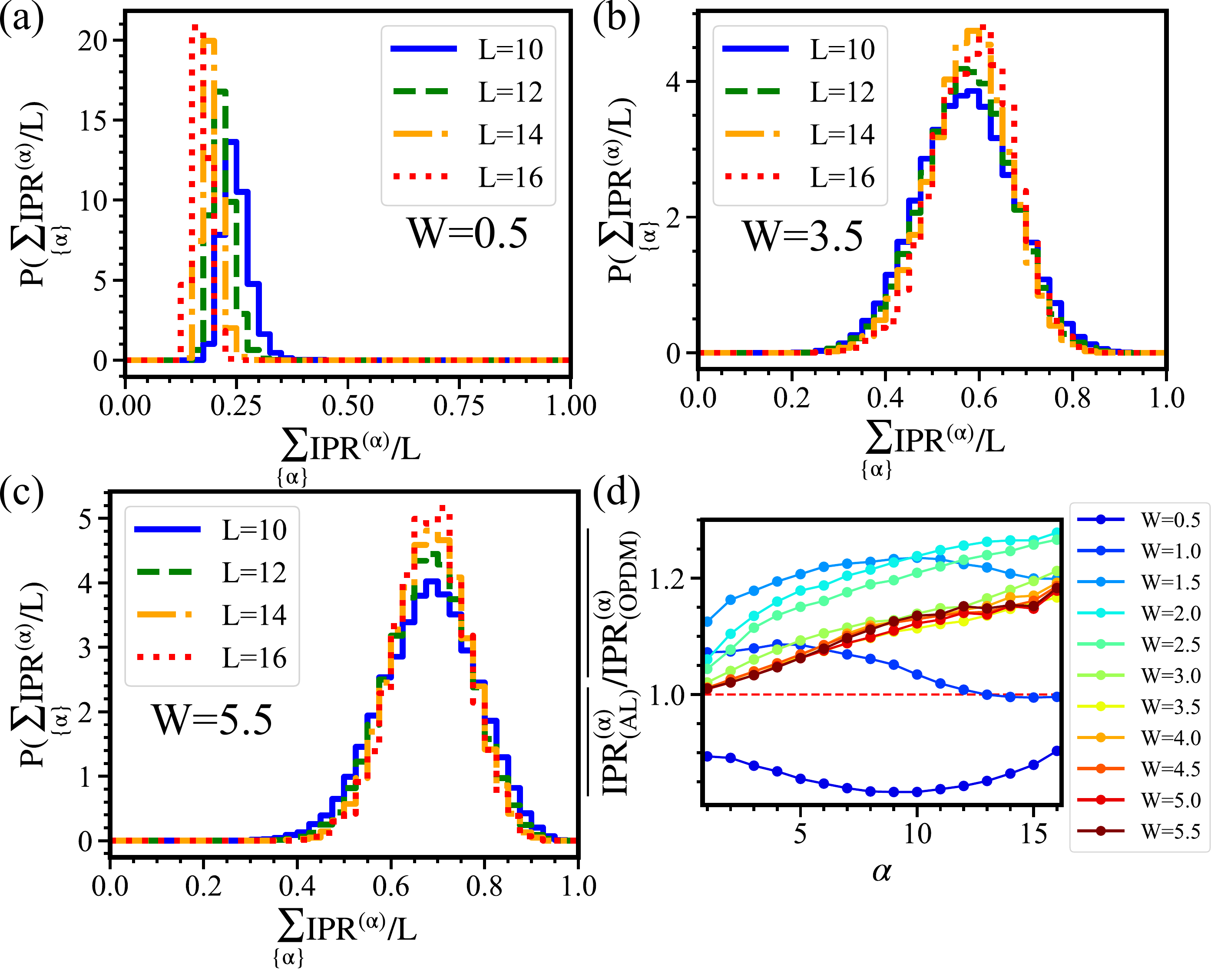}
\caption{
Probability distribution of $\sum_{\{ \alpha \}}\rm{IPR^{( \alpha)}}\rm{/L}$ for different disorder strength (a) $W=0.5$, (b) $W=3.5$, (c) $W=5.5$. (d) Ratio of IPR in AL orbital and OPDM bases.
}
\label{IPR_real_fig}
\end{figure}

\begin{figure*}
\includegraphics[width=170mm]{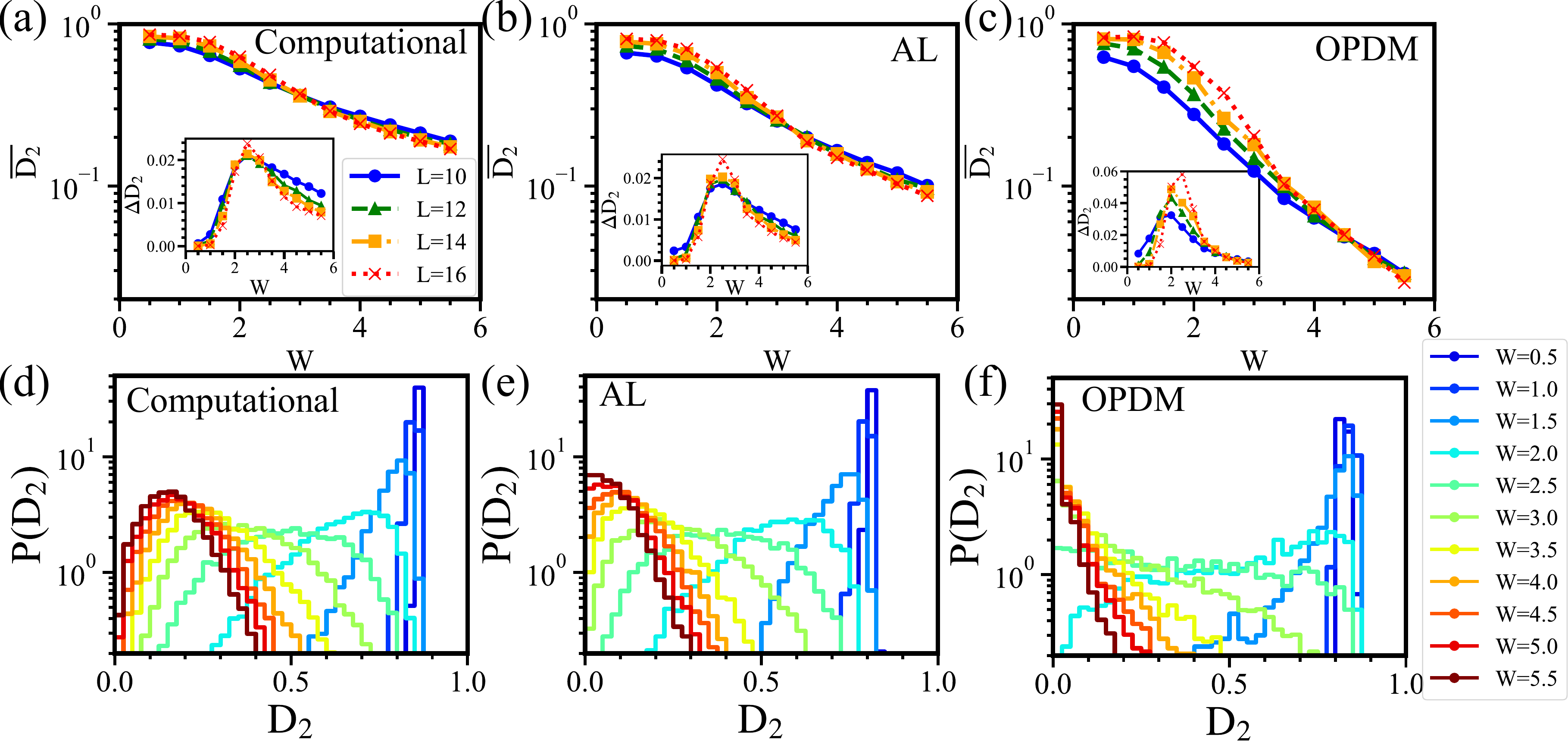}
\caption{Multifractal dimension $D_2$
[panels (a)-(c)]
and and its probability distribution
[panels (d)-(f)]
under different many-body bases
[(a), (d): the computational basis, 
(b), (e): the AL orbital basis, 
(c), (f): the OPDM basis].
The system size is $L=16$.}
\label{fig_D2}
\end{figure*}

\subsection{Natural orbitals and IPR in real space}
The eigenvectors $u^{(\alpha)}$ of the OPDM (natural orbitals),
on the other hand,
inherit the localization/delocalization nature of 
the {\it given} many-body eigenstate $|\psi \rangle$
in its spatial profile (\ref{u_alpha}).
In the MBL phase 
the ``wave function'' $\psi (j)=u^{(\alpha)}_j$
is localized exponentially in the vicinity of a localization center $j=j_\alpha$,
while it is extended in the ETH phase.\cite{Bardarson_andp}
In the non-interacting limit, 
the natural orbitals $u^{(\alpha)}_j$
reduce to the one-body 
Anderson localization orbitals
$\psi_{AL}^{(\alpha)}(j)$.

To quantify the localization/delocalization feature of the natural orbital $u^{(\alpha)}$ 
one may consider
the IPR of $u^{(\alpha)}$ in real space, i.e.,
\begin{equation}
{\rm IPR}^{(\alpha)}=\sum_{j=1}^L |u^{(\alpha)}_j|^4.
\label{IPR_real}
\end{equation}
In Fig. \ref{IPR_real_fig} (a-c)
we plot the probability distribution of IPR;
here, we have calculated
\begin{equation}
\frac{1}{L}\sum_{\{\alpha\}}{\rm IPR}^{(\alpha)}
\end{equation}
for different eigenstates $|\psi\rangle$
and for different disorder configurations,
then considered its distribution
as in the case of $P(\rho_\alpha)$ in Sec. II B
for different values of $W$
[(a) $W=0.5$, (b) $W=3.5$, (c) $W=5.5$].
The obtained $P(\sum {\rm IPR}/L)$
shows a sharply peaked distribution
in the ETH phase [case (a)]
peaked at a value $\propto 1/L$,
while
as $W$ increases,
the center of the distribution is shifted to larger values;
the peak also
broadens in the regime of ETH-MBL transition.
In the MBL phase
the center of the distribution further approaches 1.
A similar result has been reported in Fig. 3 of Ref. \onlinecite{Bardarson_PRL}.
Thus,
the real space character of the OPDM eigenvector, the natural orbitals $u^{(\alpha)}$
can be also used,
together with the distribution of its eigenvalues, the occupation spectrum,
to quantify the ETH-MBL transition.

We have repeated the same calculation in the basis of AL orbitals,
and compared the results with those in the OPDM basis.
Recall that
AL orbitals are constructed in the non-interacting limit,
while those of OPDM (natural orbitals) 
stem from an eigenstate of the full interacting system, and 
in this sense
one can naturally assume that they represent {\it dressed} AL orbitals.
The obtained
$P(\sum {\rm IPR}/L)$ in the basis of AL orbitals
shows features qualitatively similar to those in the OPDM basis,
but still differs quantitatively,
reflecting the effects of interaction in the natural orbitals.
To highlight the difference,
we plot in panel (d)
the ratio of IPR in AL and OPDM bases;
here, we have relabelled the eigenstates
in the ascending 
order of IPR such that
\begin{equation}
{\rm IPR}^{(\alpha_1)}<{\rm IPR}^{(\alpha_2)}<\cdots<{\rm IPR}^{(\alpha_L)},
\end{equation}
and consider the ratio: 
$\overline{{\rm IPR}^{(\alpha)}_{AL}}/\overline{{\rm IPR}^{(\alpha)}_{\rm OPDM}}$
at each $\alpha$,
where $\overline{\cdots}$ represents sample averaging.
In the regime of large $W$ (in the MBL phase)
the above ratio shows
a value superior to 1,
typically
for eigenstates with relatively small IPR;
i.e.,
the natural orbitals are slightly more extended than AL orbitals.
This is natural in the phenomenological LIOM picture,
since
LIOMs are considered to be dressed AL orbitals, while
the natural orbitals
are expected to mimic such LIOMs.
The eigenstates with $\alpha\simeq \alpha_L$ showing IPR$\simeq 1$ are almost frozen
and unaffected by the interaction;
the above ratio is close to 1.

\subsection{The multifractal dimension (IPR in Fock space)}

To quantify the degree of localization in the Fock space more directly, 
we here consider, instead of Eq. (\ref{IPR_real}),
the IPR in the Fock space
defined as
\begin{eqnarray}
{\cal IPR}_{\{n_j\}}&=&\sum_{\{n_j\}} \left| a_{\{n_j\}} \right|^{2q},
\label{IPR_F}
\end{eqnarray}
measuring to what extent
a many-body eigenstate $|\psi\rangle$
spreads in the Fock space spanned by 
a many-body basis as given in Eq. (\ref{basis_comp}).
To demonstrate our numerical results
we also employ a related quantity,
\begin{equation}
D_q=- \frac{\overline{\log{\cal IPR}_{\{n_j\}}}}{ \log N},
\label{mfd}
\end{equation}
called the multifractal dimension,
where $N$ is the dimension of Hilbert space 
defined in Eq. (\ref{dim_H}).
In the actual computation
we consider the typical case of $q=2$.
In Eq. (\ref{mfd}), 
we have made explicit
the specific way to take the ensemble average,
since
it may be more conventional to define 
$D_q$ such that
\begin{equation}
D_q=- \frac{\log\overline{{\cal IPR}_{\{n_j\}}}}{ \log N},
\label{mfd2}
\end{equation}
while
${\cal IPR}_{\{n_j\}}$ is not self-averaging; i.e.,
$\overline{{\cal IPR}_{\{n_j\}}}$
does not converge rapidly.\cite{santos}
Here, to accelerate this convergence,
we employ an alternative definition (\ref{mfd}), in which
we first take the logarithm of IPR$_q$ to reduce the fluctuation,
then sample average.
Note that the logarithm of the IPR$_q$ is often dubbed as 
participation R\'enyi entropy.

In Eq. (\ref{IPR_F}) we have in mind that
$|\psi\rangle$ is represented in the computational basis 
as in Eq. (\ref{psi_comp}).
In the ETH phase, 
the coefficients $a_{\{n_j\}}$'s 
are all on the same order;
i.e.,
$|a_{\{n_j\}}|\simeq 1/\sqrt{N}$, so that
IPR$_{\{n_j\}}\sim 1/N$,
or
$D_2\simeq 1$,
while
we may {\it a priori} assume that
IPR$_{\{n_j\}}\simeq 1$ in the MBL phase (at the zeroth order)
so that $D_2\simeq 0$.
Hence,
$D_2$ is presumed to show a jump: $1\rightarrow 0$
at the ETH-MBL transition.
However, as pointed out in Refs. \onlinecite{Laflo_M,Pollmann_HD},
this is not precise;
$D_2$ actually remains finite in the MBL phase.
Here, we show
through numerical simulations and the subsequent analytical considerations
to what extent this remains finite depends, however, on the basis
one employs for constructing the Fock space.

Using the OPDM creation operator (\ref{c_alpha}),
one can construct the many-body OPDM basis states:
\begin{eqnarray}
|[\alpha]\rangle_{\rm OPDM} 
&=&
c_{\alpha_{L/2}}^\dagger
\cdots 
c_{\alpha_2}^\dagger 
c_{\alpha_1}^\dagger 
|0\rangle,
\label{basis_opdm}
\end{eqnarray}
where
the notation $[\alpha]$ has been introduced
for distinguishing it from the full list $\{\alpha\}$.
Unlike $\{\alpha\}$,
$[\alpha]$ specify a selected list of states occupied in Eq. (\ref{basis_opdm}).
In terms of these OPDM basis states
we rewrite the many-body eigenstate $|\psi\rangle$ as
\begin{equation}
|\psi\rangle=\sum_{[\alpha]} a_{[\alpha]} |[\alpha]\rangle_{\rm OPDM} .
\label{psi_opdm}
\end{equation}
Using the coefficients $a_{[\alpha]}$ introduced above,
we can define the Fock-space IPR
in the OPDM basis as
\begin{equation}
{\cal IPR}_{[\alpha]} =
\sum_{[\alpha]} \left| a_{[\alpha]} \right|^4.
\label{IPR_opdm}
\end{equation}
The coefficients $a_{[\alpha]}$
are computed from those in the computational basis 
[see Eq. (\ref{a_alpha})].
Similarly, one can define the Fock-space IPR
in the AL orbital basis, employing
in Eq. (\ref{IPR_opdm})
the coefficients $a_{[\alpha]}^{(AL)}$
introduced in in Eq. (\ref{psi_AL})
instead of the $a_{[\alpha]}$'s
in Eq. (\ref{psi_opdm}).
To find these coefficients $a_{[\alpha]}$
in the natural and AL orbital bases 
is
numerically rather costly (see Appendix A).

In Fig. \ref{fig_D2}
the fractal dimension $D_2$ 
has been computed
(a) in the computational,
(b) in the AL orbital, and
(c) in the natural orbital (OPDM) bases,
and
its sample averaged values $\overline{D_2}$ are shown in the  
corresponding panels (a)-(c).
The histograms (the probability distributions) of $D_2$
corresponding to the above three panels
are shown in panels (d)-(f).
Recall that
$D_2$ is directly related to the Fock-space IPR (\ref{IPR_F})
through the relation (\ref{mfd}).

In the computational basis [panel (a) and (d)]
one can confirm the 
characteristic behavior of $D_2$
in the ETH and MBL phases reported in Refs. \onlinecite{Pollmann_HD},
i.e.,
$D_2\simeq 1$ in the ETH phase,
while
$D_2<1$ and  fluctuates
in the MBL phase.
At $W \le 1$
the histogram of $D_2$ shows a sharp peak at a value $\sim 0.8$ close 1.
In the MBL regime: $W>3.5$,
$D_2$ decreases but still takes a value $>10^{-1}$.
The histogram of $D_2$ shows a broad maximum
at a value approaching to 0 as $W$ increases.

In the OPDM basis [panel (c) and (f)]
one can see that
$D_2$ is clearly much suppressed ($<10^{-1}$) in the MBL regime
[compare the MBL regime of panel (c) and that of panel (a);
the order of $D_2$ differs; note the semi-log scale in the plots].
The shape of the histograms also differ in the MBL regime [panel (f) vs. panel (d)].
In panel (f), as $W$ increases,
the distribution $P(D_2)$ tends to be sharply peaked at $D_2=0$. 
In addition, 
the variance of $D_2$ 
shows a peak
at the ETH-MBL transition
much more enhanced in the OPDM basis; the 
height 
of the peak is three times larger ($\simeq 0.06$)
than in other
bases ($\Delta D_2\simeq 0.02$).
At the transition $W\simeq 3$
the distribution $P(D_2)$ becomes
almost uniform,
indicating that the multi-fractal dimension $D_q$ 
is actually non self-averaging.\cite{Pollmann_HD,santos}
The finiteness of the Fock-space localization length is
not only relevant to
the finiteness of $D_2$ (i.e., $D_2\neq 0$) in the MBL phase
but also to its behavior 
in the ETH-MBL 
crossover regime.
These are part of the main findings in this work.
Such peculiar behaviors of $\Delta D_2$ and $P(D_2)$ 
at the putative ETH-MBL phase transition
can be also seen away from 
the center of the spectrum: $\epsilon=0.5$, where
\begin{equation}
\epsilon=\frac{E-E_{min}}{E_{max}-E_{min}},
\label{energy_reg}
\end{equation}
with
$E_{min}$ and $E_{max}$
being respectively 
the minimum and maximum value of the eigenenergy $E$.
In Appendix C,
four panels of Fig. 8
show ETH-MBL phase diagrams
determined by the calculated values of $D_2$ and its fluctuation. 
In the OPDM basis [panel (b)]
sharply contrasting values of $D_2$ are found in the ETH and MBL phases, 
subsutantially
improving the quality of the phase diagram
as compared to the one in the computational basis [panel (a)].
In panel (d)
the fluctuation $\Delta D_2$ shows a conspicuous peak
in the ETH-MBL crossover regime,
a behavior consistent with Fig. 3 (f);
see Appendix C for details.
Results in the AL orbital basis [panel (b) and (e)] show
features intermediate between the OPDM and the computational bases.


Such a conspicuous suppression of $D_2$ 
in the MBL regime under the OPDM basis 
confirms
that
the natural orbitals 
are indeed good approximation of
the LIOM orbitals.
We have previously argued that
under the assumption that
only the first term of LIOM creation operator (\ref{LIOM})
is relevant, and
the higher order terms are negligible,
the natural orbitals $u_\alpha$
coincide with the LIOM orbitals.
We then hypothesized that
in a generic MBL situation
this assumption approximately holds,
leading to a consistent description of
the behavior of the occupation spectrum and 
its probability distribution.
Here,
the behavior of
multifractal dimension $D_2$ 
confirms this hypothesis.

In terms of the LIOM creation operators $\gamma_\alpha^\dagger$,
a many-body eigenstate $|\psi\rangle$
can be written in the simple product form as in Eq. (\ref{beta_0}). 
This means that
if $|\psi\rangle$ is represented in the (hypothetical) LIOM basis
as
\begin{equation}
|\psi\rangle=\sum_{[\beta]} a_{[\beta]} |[\beta]\rangle_{\rm LIOM},
\label{psi_LIOM3}
\end{equation}
then
a single component $a_{[\beta_0]}=1$ is finite,
and others vanish: $a_{[\alpha\neq\beta_0]}=0$.
The resulting ${\cal IPR}_{[\beta]}$ in this idealized LIOM basis is strictly 1,
and the corresponding $D_2$ strictly vanishes,
implying that
in this case
a ultimately restricted part of the total Hilbert space
is available for the realized eigenstates.
In a generic interacting many-body eigenstate $|\psi\rangle$,
such (i.e., $D_2=0$) 
usually does not happen, and is unique to the case in which
$|\psi\rangle$ is expressed as in Eq. (\ref{beta_0})
in a simple product form.

In the computational basis 
the same $|\psi\rangle$ is expressed as in Eq. (\ref{psi_comp}), or here
we rather express it as
\begin{equation}
|\psi\rangle=\sum_{[j]} a_{[j]} |[j]\rangle,
\label{psi_j}
\end{equation}
where
for specifying a basis in the computational basis
we have employed, instead of Eq. (\ref{basis_comp}),
an alternative notation:
\begin{equation}
|[j]\rangle=
c_{j_{L/2}}^\dagger
\cdots 
c_{j_2}^\dagger
c_{j_1}^\dagger 
|0\rangle.
\label{basis_j}
\end{equation}
Then, the coefficient $a_{[j]}$ is given a simple expression:
\begin{eqnarray}
a_{[j]}&=&
\langle [j]|[\beta_0]\rangle_{\rm LIOM},
\label{a_j}
\end{eqnarray}
where
$\langle [j]|[\beta]\rangle_{\rm LIOM}$
is given by the following Slater determinant:
\begin{eqnarray}
\langle [j]|[\beta]\rangle_{\rm LIOM}
&=& \det\left[
\begin{array}{cccc}
A^{(\beta_1)}_{j_1} & A^{(\beta_2)}_{j_1} & \cdots & A^{(\beta_{L/2})}_{j_1} \\
A^{(\beta_1)}_{j_2}  & A^{(\beta_2)}_{j_2}  & \cdots & A^{(\beta_{L/2})}_{j_2} \\
\vdots  & \vdots   &  &\vdots \\
A^{(\beta_1)}_{j_{L/2}} & A^{(\beta_2)}_{j_{L/2}} & \cdots & A^{(\beta_{L/2})}_{j_{L/2}} 
\end{array}
\right]
\label{slater}
\end{eqnarray}
of LIOM orbitals $A^{(\beta)}_{j}$,
which is assumed to behave as in Eq. (\ref{LIOM_exp}); i.e.,
Then,
among the coefficients $a_{[j]}$'s in Eq. (\ref{a_j}) 
the dominant contribution is from
$[j]=[j_0]$ such that
all the $j$'s in $[j_0]$
coincide
with one of the localization centers $j_\beta$ of an occupied state
in $|[\beta_0]\rangle_{\rm LIOM}$
[see Eq. (\ref{LIOM_exp})], and is found to be
\begin{eqnarray}
a_{[j_0]}&=&
\langle [j]|[\beta_0]\rangle_{\rm LIOM}
\simeq\prod_{\beta\in[\beta_0]}
A^{(\beta)}_{j_\beta}
\nonumber \\
&\simeq&
\prod_{\beta\in[\beta_0]}
\sqrt{\tanh \frac{1}{ 2\xi_\beta}}
\equiv\left(\tanh \frac{1}{ 2\overline{\xi}}\right)^{L/4},
\label{a_j0}
\end{eqnarray}
where
in the estimation of the determinants (\ref{slater}),
we have kept only the dominant contribution from the diagonal terms $A^{(\beta_j)}_{j}$,
while
in the last step we have introduced the typical localization length
$\overline{\xi}$.
The coefficients $a_{[j]}$ in the computational basis is
distributed around the single site ${[j_0]}$ in the Fock space.
At the leading order,
$a_{[j]}$ has contributions from ``sites'' ${[j']}$
such that
one of the $j$'s in $[j']$
is found to be in a neighboring site of a localization center $j_\beta$,
while
others
coincide
with one of the remaining $j_\beta$'s of an occupied state.
The contribution from such sites ${[j']}$
is on the order of
\begin{equation}
a_{[j']}\simeq e^{-1/(2\overline{\xi})},
\label{a_j1}
\end{equation}
and there are order $L$ of such sites.
As for contributions to IPR,
the dominant correction to the ideal value 1 is from
the correction given in (\ref{a_j0}),
and
the correction from $a_{[j']}$ gives only a subdominant contribution;
i.e.,
\begin{eqnarray}
IPR&\simeq& |a_{[j_0]}|^4 + L|a_{[j']}|^4 +\cdots
\nonumber \\
&\simeq&
\left(\tanh \frac{1}{ 2\overline{\xi}}\right)^{L}+Le^{-1/(2\overline{\xi})} +\cdots.
\label{IPR_a01}
\end{eqnarray}
We have checked the validity of the formulas
(\ref{a_j0}), (\ref{a_j1}) and (\ref{IPR_a01}) numerically [see Appendix B].
This, in turn,
at least indirectly certifies
the validity of our hypothesis that
the many-body eigenstate $|\psi\rangle$
can be expressed as a simple product of LIOM orbitals as in Eq. (\ref{beta_0}),
and
of the subsequent assumptions on LIOMs.

In the AL orbital basis,
one expresses $|\psi\rangle$,
instead of Eq. (\ref{psi_j}),
as
\begin{equation}
|\psi\rangle=\sum_{[\alpha]} a_{[\alpha]}^{(AL)} |[\alpha]\rangle_{AL}.
\label{psi_AL}
\end{equation}
where $|[\alpha]\rangle_{AL}$ represents the AL orbital basis:
\begin{eqnarray}
|[\alpha]\rangle_{AL} 
&=&
c_{AL}^{(\alpha_{L/2})\dagger}
\cdots
c_{AL}^{(\alpha_2)\dagger}
c_{AL}^{(\alpha_1)\dagger}
|0\rangle,
\label{basis_AL}
\end{eqnarray}
while the 
AL orbital creation operator
$c_{AL}^{(\alpha)\dagger}$ has been defined in Eq. (\ref{c_AL}).
The coefficients $a_{[\alpha]}^{(AL)}$
in Eq. (\ref{psi_AL})
are given a simple expression:
\begin{eqnarray}
a_{[\alpha]}^{(AL)} 
&=&
\ _{AL}\langle [\alpha]|[\beta_0]\rangle_{\rm LIOM},
\label{a_AL}
\end{eqnarray}
where the unitary transformation $\langle [\alpha]|[\beta]\rangle_{\rm LIOM}$
is given by the following determinant:
\begin{eqnarray}
\ _{AL}\langle [\alpha]|[\beta]\rangle_{\rm LIOM}
&=& \det\left[
\begin{array}{cccc}
\tilde{A}^{(\beta_1)}_{\alpha_1} & \tilde{A}^{(\beta_2)}_{\alpha_1} & \cdots & \tilde{A}^{(\beta_{L/2})}_{\alpha_1} \\
\tilde{A}^{(\beta_1)}_{\alpha_2}  & \tilde{A}^{(\beta_2)}_{\alpha_2}  & \cdots & \tilde{A}^{(\beta_{L/2})}_{\alpha_2} \\
\vdots  & \vdots   &  &\vdots \\
\tilde{A}^{(\beta_1)}_{\alpha_{L/2}} & \tilde{A}^{(\beta_2)}_{\alpha_{L/2}} & \cdots & \tilde{A}^{(\beta_{L/2})}_{\alpha_{L/2}} 
\end{array}
\right]
\label{slater_LIOM_AL}
\end{eqnarray}
of $\tilde{A}^{(\beta)}_\alpha$ given in Eq. (\ref{A_tilde}),
representing physically the overlap of the
$\alpha$th AL orbital and
$\beta$th LIOM orbital.
Unlike in the computational basis,
in which this overlap becomes each component of 
the LIOM wave function $A^{(\beta)}_{j}$,
here
$\tilde{A}^{(\beta)}_\alpha$
in Eq. (\ref{slater_LIOM_AL}) represents the overlap (\ref{A_tilde}),
which becomes
on the diagonal terms of Eq. (\ref{slater_LIOM_AL})
\begin{equation}
\tilde{A}^{(\beta)}_\beta
=\sum_{j} \psi_{AL}^{(\beta)*}(j) A^{(\beta)}_j,
\end{equation}
and takes a value close to 1;
at least, a value closer to 1
than the simple peak value:
\begin{equation}
A^{(\beta)}_{j_\beta}
\simeq\sqrt{\tanh\left(\frac{1}{2\xi_\beta}\right)}.
\end{equation}
One can check this hypothesis 
by assuming the following explicit form
for the AL orbitals:
\begin{equation}
\psi_\alpha^{(AL)}(j) =
\sqrt{\tanh \frac{1}{ 2\xi_\alpha^{(AL)}}}
\exp \left(-\frac{|j-j_\alpha|}{ 2\xi_\alpha^{(AL)}}\right).
\label{AL_exp}
\end{equation}
Then, one finds that the inequality
$\tilde{A}^{(\beta)}_\beta>A^{(\beta)}_{j_\beta}$
holds for a relatively broad range of $\xi_\alpha^{(AL)}$.
As a result,
${\cal IPR}_{[\alpha]}^{(AL)}$ in the AL orbital basis
dominated by the term:
\begin{eqnarray}
a_{[\beta_0]}^{(AL)}
&=&
\ _{AL}
\langle [\beta_0]|[\beta_0]\rangle_{\rm LIOM}
\simeq\prod_{\beta\in[\beta_0]}
\tilde{A}^{(\beta)}_{\beta},
\end{eqnarray}
takes a value much closer to 1 than in the computational basis,
and the corresponding $D_2$ is more suppressed
in the MBL phase.
In the OPDM basis, this tendency is further accentuated:
${\cal IPR}_{[\alpha]}\simeq 1$,
and the corresponding $D_2$ practically vanishes.
These results confirm
that
the OPDM basis is closest to LIOM orbitals;
AL orbitals are less appropriate approximation of them,
while
the effect of a finite spread of LIOM orbitals in real space [see Eq. (\ref{LIOM_exp})]
is most visible
in the Fock-space IPR represented
in the computational basis.

This observation on the practical vanishing of $D_2\simeq 0$ in the Fock space
(under the OPDM basis and in the MBL phase)
is consistent with the behavior of
the gapped occupation spectrum $\{\rho_\alpha\}$ and
the corresponding
V-shaped probability distribution $P(\rho_\alpha)$
we have described in Secs. II A and C.
In the deep MBL phase, 
the eigenstate is predominantly expressed by a product state (\ref{beta_0})
of LIOM $\simeq$ OPDM orbitals,
in which
the occupation spectrum $\rho_\alpha$
in principle
coincides with the occupation number $n_\alpha=0,1$
specifying the basis state $|[\beta_0]\rangle_{\rm LIOM}$;
$n_\alpha=1$ ($n_\alpha=0$) represents simply the occupation (vacancy)
of the $\alpha$th orbital.
The resulting probability distribution $P(\rho_\alpha)$ tends to become bimodal;
this tendency has been seen in the V-shaped distribution in Fig. \ref{fig_D2} (f).
The occupation spectrum in the computational basis is, on the other hand, 
directly susceptible of 
a finite spread of LIOM orbitals in real space [see Eq. (\ref{LIOM_exp})];
as a result 
the jump of the occupation spectrum tends to be washed out,
and the corresponding probability distribution becomes U-shaped [Fig. \ref{fig_D2} (d)].
In the AL orbital basis, we have seen features intermediate of the above two cases.





In Ref. \onlinecite{Pollmann_HD} 
it has been pointed out that
the multifractal dimension of an eigenstate 
constructed from a single Slater determinant 
remains finite and fluctuates, even in the localized phase.
This is very much consistent with what we have described above
in the case of computational basis,
and here 
it has been more thoroughly analyzed
from the scope of
our study on the basis dependence of the multifractal dimension behavior.

In reality, on the other hand,
one cannot fully escape from thermal region effects;
i.e., higher-order terms in the LIOM creation operator (\ref{LIOM})
introduce particle-hole excitations to the product state (\ref{beta_0}),
transforming it to a superposition state;
i.e., entanglement is generated. 
In real space this appears as thermal regions,
while
in Fock space
more states become available for realized eigenstates.
In the next section, 
we focus on the quantity dubbed as the local purity
and shed light on the relation
between the occupation spectrum and the multifractal dimension.

The comparison of Eqs. (\ref{IPR_F}) and Eqs. (\ref{IPR_opdm})
has been done in Ref. \onlinecite{Cheianov}, and
it was found that
the coefficients $a_{\{\alpha\}}$ in the OPDM basis
are more localized 
in the MBL phase
than $a_{\{n_j\}}$ in the computational basis;
i.e.,
IPR$_{\{\alpha\}}$ is closer to 1 than IPR$_{\{n_j\}}$
in the MBL phase;
to be precise
the authors of Ref. \onlinecite{Cheianov}
compared
the participation ratio ($=1/$IPR) in the computational and 
in the OPDM bases
[see two panels of Fig. 2 in Ref. \onlinecite{Cheianov}].

We have typically sampled $10^1-10^2$ eigenstates for each disorder realization, 
and the number of disorder realizations are $2\times10^2(W\leq2.5, L=16)-10^3$.
The total number of samples in the OPDM basis 
is more than $10^4$ for $L\leq14$ and 
$2\times10^3(W\leq2.5)-5\times10^3(W\geq3.0)$ for $L=16$. 
This is $2-5$ times more than those of Ref. \onlinecite{Cheianov}, 
and obtained results are consistent 
with those of Ref. \onlinecite{Cheianov}.

\begin{figure*}
\includegraphics[width=170mm]{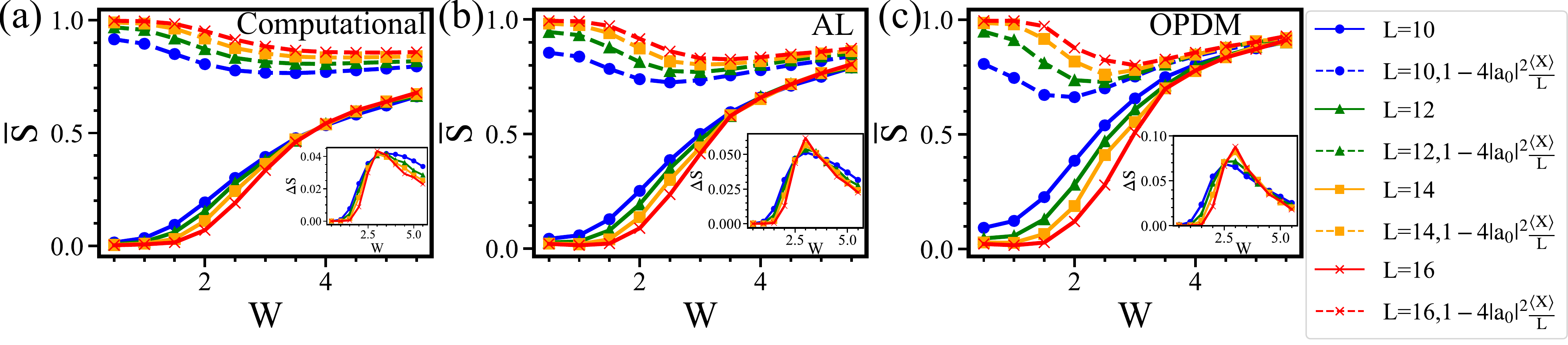}
\caption{The ensemble average of the local purity $\overline{S}$ evaluated
in different many-body bases:
(a) the computational basis,
(b) the AL orbital basis, and
(c) natural orbital (OPDM) basis,
at different strength $W$ of disorder.
In the curves represented by solid lines
$S$ is evaluated in its full form 
[as in Eq. (\ref{purity}), or in Eq. (\ref{purity_HD})], 
while in those represented by dashed lines
in an approximate form [as given in the second line of Eq. (\ref{purity_pi})].
The insets show its variance $\Delta S$.}
\label{fig_purity}
\end{figure*}

\section{Local purity and its moments, Hamming distance}

\subsection{From occupation spectrum to local purity}

The degree of Fock-space localization
is also encoded in
how close 
the occupation spectrum $\rho_\alpha$ is to a simple step function (\ref{step}).
Here, to quantify this we employ
the quantity, referred to as the local purity in Ref. \onlinecite{viola}.
The local purity $S$
of a many-body state $|\psi\rangle$ is defined as
\begin{equation}
S=\frac{1}{L}\sum_{\alpha=\alpha_1}^{\alpha_L} \langle\psi |\hat{\sigma}_{{\alpha}} |\psi\rangle^2,
\label{purity}
\end{equation}
where
\begin{equation}
\hat{\sigma}_{\alpha} = 2\hat{n}_\alpha-1 = 2c_\alpha^\dagger c_\alpha-1
\label{sigma}
\end{equation}
measures how the occupation of $j$th orbital 
$\langle\hat{n}_\alpha\rangle=\langle\psi |\hat{n}_\alpha |\psi\rangle$
deviates from 1/2;
\footnote{This is in a sense an idea presuming the situation of half-filling $\nu=N_e/L=1/2$.}
i.e.,
$\langle\hat{\sigma}_\alpha\rangle=1$ if $\langle\hat{n}_\alpha\rangle=1$
while
$\langle\hat{\sigma}_\alpha\rangle=-1$ if $\langle\hat{n}_\alpha\rangle=0$.
In a maximally localized MBL state
and if ${\alpha}_j$'s are chosen to be LIOM orbitals, then
$S=1$,
while
in the ETH phase
$\langle\hat{n}_\alpha\rangle\simeq 1/2$
so that
$S\simeq 0$.
Here, we have in mind a situation
in which
the state $|\psi\rangle$ is expressed as a superposition of
many-body basis states $|\mu\rangle$ as
\begin{equation}
|\psi\rangle = \sum_{\mu} a_\mu |\mu\rangle,
\label{psi_mu}
\end{equation}
where
$\mu$ is, {\it e.g.}, specified by a Fock representation:
\begin{eqnarray}
\mu=\{n_\alpha\}=(n_{\alpha_1}, n_{\alpha_2}, \cdots, n_{\alpha_L}).
\label{mu_n}
\end{eqnarray}
The integers
$n_\alpha=0,1$
specify
how many electrons are in the state $\alpha$ ($\alpha$th orbital)
in the basis state $|\mu\rangle$.
For basis orbitals $\alpha$'s
we consider the specific cases of natural (OPDM) and AL orbitals
as concrete examples.
In the case of computational basis
a basis state $|\mu\rangle$ specified by Eq. (\ref{mu_n}) 
reduces to Eq. (\ref{basis_comp}).
Note that
in the new notation $\mu$
we label a many-body basis state 
such as the ones given in Eqs. (\ref{basis_comp}), (\ref{basis_opdm}), or (\ref{basis_AL})
by a single label $\mu$, which is later specified by a single number:
$\mu=0,1,2,\cdots,N-1$.
Note also that the choice of our operator
$\hat{\sigma}_\alpha$
in Eq. (\ref{sigma})
is basis dependent, and chosen in such a way that
$\hat{\sigma}_\alpha$ is a good quantum number for a basis state $|\mu\rangle$;
i.e.,
$\hat{\sigma}_\alpha=+1$ if $n_\alpha=1$ in $\mu$, while
$\hat{\sigma}_\alpha=-1$ if $n_\alpha=0$ in $\mu$.
Thus, the contribution of a basis state $|\mu\rangle$
to the quantity
$\langle\hat{\sigma}_\alpha\rangle=\langle\psi |\hat{\sigma}_\alpha |\psi\rangle$
is either $+1$ if $n_\alpha=1$ in $\mu$,
or $-1$ if $n_\alpha=0$ in $\mu$.
Based on these two types of contributions to $\langle\hat{\sigma}_\alpha\rangle$,
let us classify the set of $N$ basis states $\{\mu\}$ into two categories:
$\{\alpha_\uparrow\}$ and $\{\alpha_\downarrow\}$,
where
$\mu\in\{\alpha_\uparrow\}$ if $n_\alpha$ in Eq. (\ref{mu_n}) is $+1$, while
$\mu\in\{\alpha_\downarrow\}$ if $n_\alpha$ in Eq. (\ref{mu_n}) is $-1$.
This allows us to express 
$\langle\hat{\sigma}_\alpha\rangle$
as
\begin{eqnarray}
\langle\hat{\sigma}_\alpha\rangle&=&\langle\psi |\hat{\sigma}_\alpha |\psi\rangle
\nonumber \\
&=&
\left(
\sum_{\mu\in\{\alpha_\uparrow\}} - \sum_{\mu\in\{\alpha_\downarrow\}}
\right)
|a_\mu|^2
\label{sigma_j}
\end{eqnarray}
Substituting Eq. (\ref{sigma_j}) into the definition of the local purity (\ref{purity}),
one can prove \cite{viola}
(see Appendix B),
\begin{eqnarray}
S&=&\frac{1}{L}\sum_{\alpha=\alpha_1}^{\alpha_L} \langle \hat{\sigma}_\alpha\rangle^2
\nonumber \\
&=&
1-
\frac{4}{L}\sum_{\mu<\nu}
x_{\mu\nu}
|a_\mu|^2
|a_\nu|^2,
\label{purity2}
\end{eqnarray}
where
$x_{\mu\nu}$ represents the Hamming distance between the two basis states
$|\mu\rangle$ and $|\nu\rangle$.
In the last step 
to find the final expression for the purity in Eq. (\ref{purity2})
we have noted the following identity:
\begin{equation}
\sum_{\alpha=\alpha_1}^{\alpha_L}
\sum_{\mu\in\{\alpha_\uparrow\}}\sum_{\nu\in\{\alpha_\downarrow\}}
|a_\mu|^2
|a_\nu|^2
=\sum_{\mu<\nu}x_{\mu\nu}
|a_\mu|^2
|a_\nu|^2
\label{sum_ham}
\end{equation}
where
$\sum_{\mu\in\{\alpha_\uparrow\}}$,
$\sum_{\nu\in\{\alpha_\downarrow\}}$,
or rather the underlying classification of the states $|\mu\rangle$ into
$\{\alpha_\uparrow\}$ and $\{\alpha_\downarrow\}$
is (implicitly) dependent on $j$,
since at each $j$ we have classified $|\mu\rangle$ 
based on the value of $n_\alpha$ in $|\mu\rangle$.
To verify Eq. (\ref{sum_ham})
let us consider
how many times
a given combination 
$|a_\mu|^2 |a_\nu|^2$
appears in the summation on the l.h.s. of Eq. (\ref{sum_ham}).
Provided that $\mu\neq\nu$,
$n_\alpha$'s in $\mu$ and $\nu$ differ at least some $\alpha$,
and at this $\alpha\equiv\tilde{\alpha}_1$
either
$\mu\in\{\alpha_\uparrow\}$ and $\nu\in\{\alpha_\downarrow\}$
or vice versa
holds;
i.e.,
at this $\alpha$
the combination $|a_\mu|^2 |a_\nu|^2$
is eligible for being included
in the summation on the l.h.s. of Eq. (\ref{sum_ham}),
i.e., in
$\sum_{\mu\in\{\alpha_\uparrow\}}\sum_{\nu\in\{\alpha_\downarrow\}}|a_\mu|^2 |a_\nu|^2$.
If $n_\alpha$'s in $\mu$ and $\nu$ also differ at some another
$\alpha=\tilde{\alpha}_2$,
then
at this value of $\alpha=\tilde{\alpha}_2$
the combination $|a_\mu|^2 |a_\nu|^2$
is again eligible for being included
in the same summation;
and the same may happen again at $\alpha=\tilde{\alpha}_3$, etc.
Thus,
the combination $|a_\mu|^2 |a_\nu|^2$
appears 
on the l.h.s. of Eq. (\ref{sum_ham})
$x_{\mu\nu}$ times,
where
$x_{\mu\nu}$ is
the number of times in which
$n_\alpha$'s in $\mu$ and $\nu$ differ,
and this number
$x_{\mu\nu}$,
measuring the distance in Fock space
between the two states
$|\mu\rangle$ and $|\nu\rangle$
is often referred to as Hamming distance.
\cite{Pollmann_HD,Logan2019,Logan2020}

In Fig. \ref{fig_purity},
the purity $S$
evaluated in the three different many-body bases;
(a) computational,
(b) AL orbital, and
(c) natural orbital (OPDM) bases.
is shown as a function of $W$
and compared.
In the three bases,
one can see the expected tendencies:
i.e.,
the purity $S$
tends to vanish in the ETH phase:
$S\simeq 0$
as $L$ increase,
while
in the MBL phase
$S$
takes a value on the order of unity.
Thus, in the limit of large $L$,
as $W$ increases,
the value
$S=0$ in the ETH phase
is expected to jump into a finite value
at the ETH-MBL transition: $W\simeq 3.5-4$,
while
the magnitude of this jump
$\delta S$
is largest (smallest) in the OPDM (computational) basis.
As $W$ is further increased,
the value of $S$ tends to approach the ideal value 1
in the OPDM basis,
while
in the computational basis
it remains to be a value considerably smaller than 1.
In the AL orbital basis, features intermediate between the two cases
are seen.

In the computational basis
the deviation of
$S$ from the ideal value 1
has two independent 
sources.
One is the imperfection of the Fock-space localization in the MBL phase
due to higher-order terms in Eq. (\ref{LIOM}); i.e., the effect of thermal regions,
while the other is
the finite localization length; i.e., a finite spread of the hypothetical LIOM wave functions
in real space
[cf. Eq. (\ref{LIOM_exp})].
We have previously seen that
in the computational basis
one is strongly susceptible of the second effect;
cf. discussion on the multifractal dimension in the computational basis (Sec. II-E).
On the other hand,
in the OPDM basis,
one is almost free from this extrinsic effect (second effect).
Therefore,
the remaining deviation of
$S$ in the MBL phase from the ideal value 1
in the OPDM basis
quantifies
the degree of intrinsic imperfection of the Fock-space localization 
due to the presence of thermal regions,
and 
the finiteness of the jump $\Delta S$
is possibly related to the KT nature of the
ETH-MBL transition.

\begin{figure*}
\includegraphics[width=170mm]{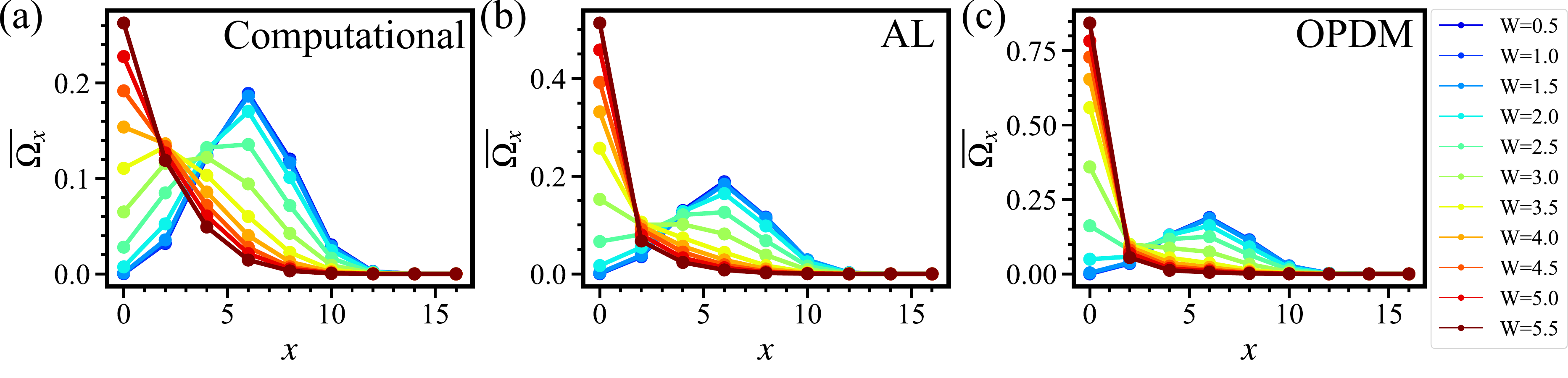}
\caption{The distribution $\Omega_x$ [see Eq. (\ref{Ax})] of the Hamming distance $x$
computed
in the different many-body bases:
(a) computational,
(b) AL orbital, and
(c) natural orbital (OPDM).
}
\label{Ax_x}
\end{figure*}

\subsection{Purity vs. Fock-space IPR}
Let us compare the expression (\ref{purity2}) for the local purity 
with the one for IPR in Fock space:
\begin{eqnarray}
{\cal IPR} &=&
\sum_\mu
|a_\mu|^4
\nonumber \\
&=&
1-2\sum_{\mu<\nu} 
|a_\mu|^2
|a_\nu|^2.
\label{IPR_F2}
\end{eqnarray}
In the second line,
${\cal IPR}$
has been rewritten in a form similar to Eq. (\ref{purity2});
see Appendix C for its derivation.
The local purity $S$ and ${\cal IPR}$ are similar quantities, both
measuring the degree of Fock-space localization,
taking values $\simeq 1$ in the MBL (Fock-space localized) phase, while
$\simeq 0$ in the ETH (Fock-space delocalized) phase. 
Comparing Eqs. (\ref{purity2}) and (\ref{IPR_F2}),
one can see that the two quantities, indeed follow a similar expression,
except one remarkable difference;
in the sum collecting the contributions from pairwise amplitudes $|a_\mu|^2|a_\nu|^2$,
each contribution is weighted in Eq. (\ref{purity2})
by the Hamming distance $x_{\mu\nu}$ between the two basis states 
between
$|\mu\rangle$ and $|\nu\rangle$,
while it is not in Eq. (\ref{IPR_F2}).

To quantify the contributions from different Hamming distance pairs
let us introduce the quantity:
\begin{equation}
\Omega_x=\sum_{\mu<\nu s.t. x_{\mu\nu}=x} |a_\mu|^2|a_\nu|^2,
\label{Ax}
\end{equation}
where 
the summation is for all pairs
$\mu<\nu$
such that
their Hamming distance $x_{\mu\nu}$ is
constrained to a specific value $x=1,2,\cdots$.
For $x=0$, let us define $\Omega_0$ such that
\begin{equation}
\Omega_0=\sum_{\mu} |a_\mu|^4={\cal IPR}_{\{a_\mu\}}
\label{A0}
\end{equation}
Using $\Omega_x$, one can reexpress the second term of Eq. (\ref{purity2}) as
\begin{eqnarray}
\sum_{\mu<\nu}x_{\mu\nu}
|a_\mu|^2
|a_\nu|^2
&=&
\sum_x x \sum_{\mu<\nu s.t. x_{\mu\nu}=x} |a_\mu|^2|a_\nu|^2
\nonumber \\
&=&
\sum_{x=1}^\infty x \Omega_x,
\end{eqnarray}
i.e.,
\begin{eqnarray}
S&=&
1-\frac{4}{L}\sum_{x=1}^\infty
x \Omega_x.
\label{purity_HD}
\end{eqnarray}
Using $\Omega_x$, one can also rewrite Eq. (\ref{IPR_F}) as
\begin{eqnarray}
{\cal IPR} &=&
1-2\sum_{x=1}^\infty
\Omega_x.
\label{IPR_HD}
\end{eqnarray}
The form of
Eqs. (\ref{purity_HD}) and (\ref{IPR_HD})
suggests that
$\Omega_x$ plays the role of a probability distribution for $x$.
It indeed is for the occurrence of a pair $c_{\alpha_\mu}^\dagger$ and $c_{\alpha_\nu}^\dagger$
in the given many-body (eigen)states $|\psi\rangle$.
Also,
the summation in
Eqs. (\ref{purity_HD}) and (\ref{IPR_HD})
starts from $x=1$ in the generic case, while
in our present context
with a fixed electron density (to $\nu=1/2$),
the summation actually starts from 2,
and
$x$ takes only even integer values provided that $L$ is even.

In three panels of Fig. \ref{Ax_x}
the pairwise distribution
$\Omega_x$ 
computed under different bases
has been ensemble averaged
and plotted against the Hamming distance $x$.
The plots show the evolution of the
distribution $\Omega_x$ as a function of $W$.
At weak $W$; i.e. in the ETH phase,
the distribution $\Omega_x$
shows a broad maximum
around
the center of the Fock-space hypercube
$x\simeq L/2$,
while as $W$ increases,
the center of mass of the distribution
gradually shifts to $x=0$, 
and in the MBL phase
$\Omega_x$ becomes maximal at $x=0$.
In the OPDM basis [panel (c)]
$\Omega_x$ is sharply peaked at $x=0$;
the peak is sharp and high: $\Omega_0\simeq 0.8$,
while
in the computational basis [panel (a)]
$\Omega_x$ exhibits a relatively long tail
that extends almost as far as $x=L/2$.
In the AL orbital basis
the behavior of $\Omega_x$ is intermediate between the two cases.

The height of the peak at $x=0$ is
$\Omega_0$ [see Eq. (\ref{A0})], 
which is nothing but the value of Fock-space IPR
discussed in the previous section.
In the computational basis [panel (a)]
the Fock-space IPR is, as was also the case in the local purity $S$, 
strongly susceptible of
the finite localization length; i.e., a finite spread of the hypothetical LIOM wave functions
in real space
[cf. Eq. (\ref{LIOM_exp}) and Eqs. (\ref{a_j0}), (\ref{a_j1}), (\ref{IPR_a01})].
Note that the peak of the LIOM wave function diminishes since it spreads;
diminishes to compensate for the normalization,
giving the principal reason why
$\Omega_0$ is relatively small in panel (a).
In the OPDM basis [panel (c)], on the other hand,
this extrinsic effect is almost removed
so that
the peak height $\Omega_0$ is close to the ideal value 1
in the deep MBL regime.
To quantify the behavior of $\Omega_x$ in a broader regime of $W$,
let us further analyze the nature of this quantity
in the next subsection.




\begin{figure*}
\includegraphics[width=170mm]{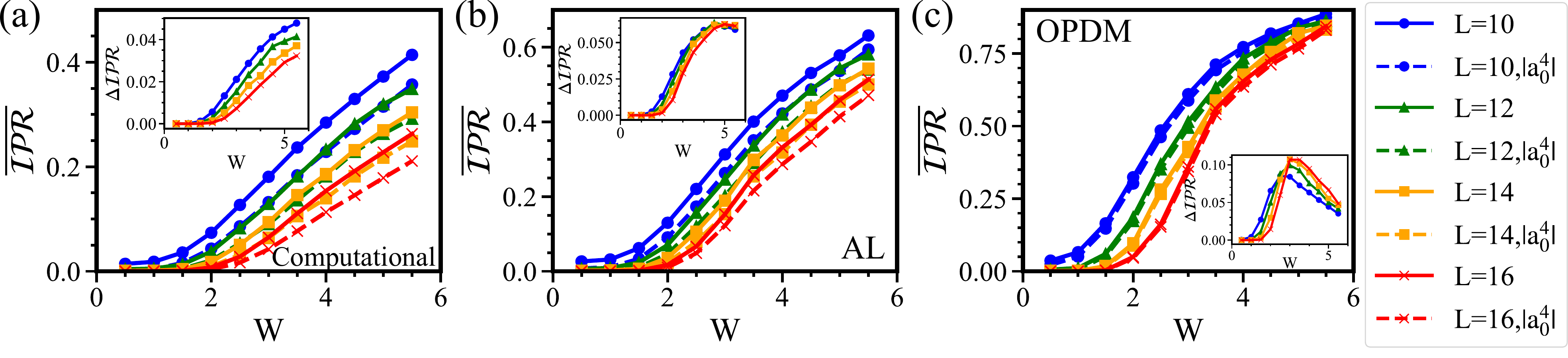}
\caption{
Comparison of the total Fock-space IPR and 
the contribution from $|a_0|^4$ [cf. Eq. (\ref{IPR_pi})]
(a) in the computational,
(b) in the AL orbital, and
(c) in the natural orbital (OPDM)
bases.
The insets show the variance $\Delta {\cal IPR}$.
}
\label{fig_IPR_asymp}
\end{figure*}

\subsection{$\Omega_x$ vs. $\Pi(x)$}
In the MBL phase
the state $|\psi\rangle$ is close to a simple product 
state
such as the one given in Eq. (\ref{beta_0}); i.e.,
in the superposition of Eq. (\ref{psi_mu}),
the contribution from a single component $\mu=\mu_0$ predominates,
and 
those from other basis states $\mu\neq\mu_0$ are minor.
To make clear the situation,
let us relabel $\mu$'s such that
\begin{equation}
|a_0| > |a_1| > |a_2| >\cdots,
\end{equation}
then one can naturally assume 
that
\begin{equation}
|a_0| \gg |a_1|, |a_2|, \cdots.
\label{a0>>}
\end{equation} 
In Fig. \ref{Ax_x} (c)
the value of $\Omega_0$ [cf. Eq. (\ref{A0})]
is closest to 1 in panel (c); {\it i.e.}, in the OPDM basis:
$\Omega_0= |a_0|^4 \simeq 0.8$
in the deep MBL regime,
so that $|a_0|^2$ will be even closer to 1.
Therefore, $|a_0|^2 \gg$ all other $a_\mu$'s.
The inequality (\ref{a0>>})
signifies
that among the contribution from various pairs $|a_\mu|^2|a_\nu|^2$
in Eq. (\ref{Ax})
only terms containing $|a_0|^2$ give principal contributions, i.e.,
Eq. (\ref{Ax}) can be well-approximated as 
\begin{eqnarray}
\Omega_x&=&\sum_{\mu<\nu s.t. x_{\mu\nu}=x} |a_\mu|^2|a_\nu|^2
\nonumber \\
&\simeq&
|a_0|^2
\sum_{\mu {\rm s.t.} x_{\mu,0}=x}|a_\mu|^2
\equiv
|a_0|^2\Pi (x).
\label{Ax2}
\end{eqnarray}
In the last line we have introduced the quantity
\begin{equation}
\Pi (x)=\sum_{\mu {\rm s.t.} x_{\mu,0}=x}|a_\mu|^2,
\label{radial_D}
\end{equation}
which has been dubbed as {\it radial distribution} in Ref. \onlinecite{Pollmann_HD}.
As a distribution,
this quantity $\Pi(x)$ may be better behaved than our $\Omega_x$ given in Eq. (\ref{Ax})
in the sense that
with a natural convention of $\Pi (0)=|a_0|^2$,
$\Pi(x)$ has been automatically normalized:
\begin{eqnarray}
\sum_{x=0}^\infty \Pi (x)&=&\sum_{x=0}^\infty \sum_{\mu {\rm s.t.} x_{\mu,0}=x}|a_\mu|^2
=\sum_{\mu}|a_\mu|^2=1.
\end{eqnarray}
The radial distribution $\Pi(x)$ 
measures the Hamming distance $x$ from a principal component $|\mu_0\rangle$,
while
the distribution
$\Omega_x$ 
measures its occurrence
in the entire distribution of basis states $|\mu\rangle$.
In the light of Eq. (\ref{Ax2})
let us further interpret the results shown in Fig. \ref{Ax_x}.
We have previously considered the case
in which $W$ is strong enough for the system to be in the deep MBL limit,
where
the many-body eigenstate $|\psi\rangle$
is expressed by a simple product state as in Eq. (\ref{beta_0}),
then
$\Omega_x$ is sharply peaked at $x=0$.
This is typically the case in panel (c) in the OPDM basis, since
the natural orbitals $\simeq$ LIOMs, and
in the OPDM basis
$|\psi\rangle$ may still be well approximated by the simple product state:
\begin{equation}
|\psi\rangle\simeq
c_{\beta_{L/2}}^\dagger
\cdots 
c_{\beta_2}^\dagger
c_{\beta_1}^\dagger 
|0\rangle
\equiv |[\beta_0]\rangle_{\rm OPDM},
\label{opdm_0}
\end{equation}
where
$c_{\beta}^\dagger$ creates an electron in the $\beta$th natural orbital.
As $W$ decreases, however,
higher-order terms of LIOM; i.e., such as the $B$-terms in Eq. (\ref{LIOM}) 
become non-negligible, 
and add to the principal product state (\ref{opdm_0})
those terms that can be created by particle-hole excitations;
i.e.,
\begin{equation}
|\psi\rangle\simeq
\left[
1+\sum_{\alpha\beta} \tilde{B}_{\alpha\beta}c_{\alpha}^\dagger c_{\beta}+\cdots
\right]
|[\beta_0]\rangle_{\rm OPDM},
\label{opdm_01}
\end{equation}
where
the second and higher-order terms create states
that are detached from $|[\beta_0]\rangle_{\rm OPDM}$ in Fock space
by the Hamming distance $x$ equal to twice the number of particle-hole excitations.
As $W$ decreases and the system
approaches the MBL-ETH transition,
such higher-order terms tend to become more important.
As a result, the state $|\psi\rangle$ initially point-localized in the Fock space
at $[\beta]=[\beta_0]$
acquires a finite expanse
specified by the distribution $\Omega_x$.
In panels (c) and (b) of Fig. \ref{Ax_x}
the distribution 
$\Omega_x$ 
is sharply peaked at $x=0$
in the deep MBL phase, while
as $W$ decreases
it resolves itself into a broader distribution
extended to the region of $x\neq 0$.
Such an evolution is well explained by the appearance of
higher-order terms in Eq. (\ref{opdm_01})
that physically
represent particle-hole excitations.
How much weight
the distribution $\Omega_x$ has away from $x=0$
is a measure of
to what degree
such states created by particle-hole excitations are mixed
with the principal product state $|[\beta_0]\rangle_{\rm OPDM}$
in the realized eigenstate $|\psi\rangle$.

In terms of $\Pi(x)$ and in the MBL phase
Eq. (\ref{purity_HD})
may be rewritten as
\begin{eqnarray}
S&=&1-\frac{4}{L}\sum_{x=1}^\infty x\Omega_x
\nonumber \\
&\simeq&
1-\frac{4}{L}|a_0|^2
\sum_{x=1}^\infty
x\Pi(x).
\label{purity_pi}
\end{eqnarray}
In Fig. \ref{fig_purity}
the purity $S$ 
is evaluated
both in its full [as in the first line of Eq. \ref{purity_pi})]
and asymptotic [as in the second line of Eq. \ref{purity_pi})]
forms,
and they are
plotted together for comparison
in the three different bases:
(a) computational,
(b) AL orbital, and
(c) natural orbital (OPDM).
The two quantities 
tend to merge in the MBL phase
in the three bases, but
the agreement is best and almost perfect
in the OPDM basis
and
in the deep MBL regime.
Note that
the two quantities coincide
signifies that
the principal component $a_0$ is indeed predominant,
and the assumption (\ref{a0>>})
is well justified.
The fact that
$S$ deviates significantly from its asymptotic expression [the second line of Eq. \ref{purity_pi})]
signifies that
the quantity $S$ well describes
the expanse the weight of $|\psi\rangle$ on $a_\mu$ around $a_0$.

Using $\Omega_x$, one can also rewrite Eq. (\ref{IPR_HD}) as
\begin{eqnarray}
{\cal IPR} &=&1-2\sum_{x=1}^\infty \Omega_x
\nonumber \\
&\simeq&
1-2|a_0|^2
\sum_{x=1}^\infty \Pi(x)
\nonumber \\
&=&
1-2|a_0|^2+2|a_0|^4\ge |a_0|^4.
\label{IPR_pi}
\end{eqnarray}
In Fig. \ref{fig_IPR_asymp}
the total Fock-space IPR and 
the contribution from $|a_0|^4$ (cf. the last expression above)
have been plotted together and compared
in the three different bases:
(a) computational,
(b) AL orbital, and
(c) natural orbital (OPDM).
The plots show that
actually
in all the bases
and in all range of $W$,
the two quantities almost coincide; i.e.,
$|a_0|^4$ is a good approximation of the Fock-space IPR,
and
the agreement is almost perfect
in the OPDM basis.
This, in turn,
implies that
unlike the local purity $S$ in Fig. \ref{fig_purity}
the Fock-space IPR
is almost 
exclusively determined by the principal term $a_0$
and 
not much sensitive to the expanse 
of the weight of $|\psi\rangle$ in the Fock space around $a_0$.

\section{Concluding remarks}
To highlight the nature of
many-body localization (MBL),
especially 
focusing on its aspect of
Fock-space localization,
we have considered a paradigmatic model of MBL;
a one-dimension spinless fermion model (\ref{ham}).
As a practical tool of the analysis
we have employed the one-particle density matrix (OPDM) approach [see Eq. (\ref{rho_ij})].
The natural orbitals, i.e., the eigenvectors of the OPDM are 
expected to mimic
the local integrals of motion (LIOMs)
emergent in the MBL phase.
We have thus expected 
that
the use of natural orbitals as basis states (i.e., the use of OPDM basis),
minimizing effects of the finiteness of Fock-space localization length,
qualitatively improves
our description 
of the ETH-MBL crossover regime.

We begin by
investigating the occupation spectrum $\rho_\alpha$ (the eigenvalues of OPDM)
and the natural orbitals
as a measure
for quantifying the degree of Fock-space localization
prevailing in the system.
In the MBL phase, 
$\rho_\alpha$
becomes almost bimodal, taking values close to either 0 or 1,
implying that
the predominant part of the eigenstate $|\psi\rangle$ 
is expressed by a simple product of basis orbitals 
[Eq. (\ref{beta_0})]. 
In other words,
the system is strongly Fock-space localized.
To visualize this situation 
the distribution of $\{\rho_\alpha\}$ has been represented
in the form of a sharp step function [Fig. \ref{occ_spec} (c)].
In the computational and AL orbital bases
the occupation spectrum exhibits much smeared-off steps [see panels (a) and (b) of Fig. \ref{occ_spec}],
implying that
$\rho_\alpha$ fluctuates strongly [see U-shaped distribution of $P(\rho_\alpha)$ in panels (a) and (b) of Fig. 2].
We have shown that
this strong fluctuation of $\rho_\alpha$ in the computational and AL orbital bases
stems from 
a finite spread; i.e., a finite localization length
of the LIOM wave functions 
in real space [Eq. (\ref{LIOM_exp})].

We have also investigated the multifractal dimension $D_q$
in the computational, AL orbital and OPDM bases.
It has been previously suggested
that $D_q$ 
is also affected by the fluctuations due to a finite localization length,
alike in the case of occupation spectrum. 
In the OPDM basis 
$D_q$ exhibits a conspicuously strong suppression [Fig. 3 (c)]
in the MBL phase
and
it also weakly fluctuates [see Fig. 3 (f)].
These imply that
the natural orbitals are good approximation of LIOM orbitals,
and correspondingly,
quantities represented in the OPDM basis
are immune to
extrinsic fluctuations 
induced by a finite localization length of the LIOM orbitals.
Thus the use of OPDM basis, 
indeed minimizing the effects of the finiteness of Fock-space localization length,
improves our description of the 
ETH-MBL 
crossover regime;
see also the phase diagram in Appendix C.
Our analysis
shows that the finiteness of $D_q$ in the computational basis
reported in the literature
is indeed due to the finiteness of the Fock-space localization length.

Finally, we have introduced the quantity $\Omega_x$ [Eq. (\ref{Ax})]
which plays the role of linking 
the Fock-space IPR $\simeq$ multifractal dimension 
with the local purity,
an index quantifying the nature of occupation spectrum. 
This 
$\Omega_x$
characterizes how the many-body wave function $|\psi\rangle$ spreads
in the Fock space,
using the Hamming distance
as the metric in this space.
In the ETH phase, 
$\Omega_x$
shows a broad maximum around the center of the Fock-space hypercube $x=L/2$, 
while in the MBL phase, 
it is peaked at $x=0$.
The center of mass of the distribution $\Omega_x$ 
is directly linked to the local purity,
while
$\Omega_0$ represents the Fock-space IPR. 
The departure of $\Omega_x$ from $x=0$;
i.e., the departure of weight of $|\psi\rangle$ from the principal component $a_0$
is identified as
contributions from particle-hole excitations [see Eq. (\ref{opdm_01})],
stemming from
non-negligible higher-order corrections 
in the LIOM
creation operator (\ref{LIOM});
note that such terms must appear in the perturbative expansion of LIOM.
Interestingly,
the lower bound of the Fock-space IPR and of the local purity 
is both given
in terms of $a_0$ [see Eqs. (\ref{purity_pi}) and (\ref{IPR_pi})].
In particular, it is interesting to note that in the OPDM basis, the Fock-space IPR 
is well approximated by $a_0^4$ 
implying that when a finite jump occurs in $D_q$ 
it is also expected to occur in the local purity.

\acknowledgments
We are indebted to QuSpin\cite{Quspin1,Quspin2}
for fascilitating
the diagonalization of a many-body Hamiltonian such as the one given in Eq. (1).
K.I. is supported by JSPS KAKENHI Grant Number
21H01005,
20K03788 and 18H03683.

\appendix
\section{Unitary transformation of the many-body basis}

To compute, {\it e.g.},
the Fock-space IPR (\ref{IPR_F})
in the computational basis,
one needs to find the coefficients $a_{\{n_j\}}$ 
in Eq. (\ref{psi_comp}).
For that,
it suffices to
{\it once} diagonalize
(numerically) 
the many-body Hamiltonian (\ref{ham}); i.e.,
the many-body eigenstate $|\psi\rangle$
is an eigenvector of the $N\times N$ matrix
\begin{equation}
H_{\{m_i\}\{n_j\}}=\langle\{m_i\} | H |\{n_j\}\rangle,
\label{H_mn}
\end{equation}
where to be precise, Eq. (\ref{H_mn}) gives its $(\{m_i\},\{n_j\})$-matrix element,
and
the coefficient $a_{\{n_j\}}$ is the $\{n_j\}$th component of the eigenvector $|\psi\rangle$.
To compute
the same quantity
in a localized orbital basis,
one needs to find the coefficients $a_{[\alpha]}$
as given in Eq. (\ref{psi_opdm}),
using a unitary transformation, 
from the ones in the computational basis.
Numerically, this turns out to be rather costly. 

\subsection{OPDM (natural orbital) basis}

Noticing the completeness of the OPDM basis,
\begin{equation}
1=\sum_{[\alpha]} |[\alpha]\rangle \langle [\alpha]|,
\end{equation}
one can rewrite Eq. (\ref{psi_comp}) as
\begin{eqnarray}
|\psi\rangle &=& \sum_{\{n_j\}} a_{\{n_j\}} |\{n_j\}\rangle
=\sum_{[j]} a_{[j]} |[j]\rangle
\nonumber \\
&=& \sum_{[j]} a_{[j]} \sum_{[\alpha]} |[\alpha]\rangle \langle [\alpha]|[j]\rangle
\nonumber \\
&=& \sum_{[\alpha]}
\left(\sum_{[j]} a_{[j]} \langle [\alpha]|[j]\rangle
\right)
|[\alpha]\rangle
\nonumber \\
&\equiv&\sum_{[\alpha]} a_{[\alpha]}^{\rm (OPDM)} |[\alpha]\rangle_{\rm OPDM},
\end{eqnarray}
where in the first line,
we have rewritten the Fock representation $\{n_j\}$
in a different notation (\ref{a_j}),
which is more convenient here.
The last identity 
signifies that
the the coefficients $a_{[\alpha]}$ in the opdm basis can be computed from $a_{\{n_j\}}$
or from $a_{[j]}$,
using the relation:
\begin{equation}
a_{[\alpha]}^{\rm (OPDM)} =\sum_{[j]} a_{[j]} \ _{\rm OPDM}\langle [\alpha]|[j]\rangle,
\label{a_alpha}
\end{equation}
where
$\ _{\rm OPDM}\langle [\alpha]|[j]\rangle=\langle [j]|[\alpha]\rangle_{\rm OPDM}^*$
can be calculated from the following Slater determinant:
\begin{eqnarray}
\langle [j]|[\alpha]\rangle_{\rm OPDM} = \det\left[
\begin{array}{cccc}
u^{(\alpha_1)*}_{j_1} & u^{(\alpha_2)*}_{j_1} & \cdots & u^{(\alpha_{L/2})*}_{j_1} \\
u^{(\alpha_1)*}_{j_2}  & u^{(\alpha_2)*}_{j_2}  & \cdots & u^{(\alpha_{L/2})*}_{j_2} \\
\vdots  & \vdots   &  &\vdots \\
u^{(\alpha_1)*}_{j_{L/2}} & u^{(\alpha_2)*}_{j_{L/2}} & \cdots & u^{(\alpha_{L/2})*}_{j_{L/2}} 
\end{array}
\right].
\label{slater_opdm}
\end{eqnarray}
Let us finally estimate
how costly is to calculate the coefficients $\{a_{[\alpha]}\}$;
i.e., the set of $N$ coefficinets
in the OPDM basis
from the ones in the computational basis
through
the unitary transformation (\ref{a_alpha}).
The complexity of this task may be estimated as
\begin{equation}
N^2\times (L/2)^4 \times ({\rm  number\ of\ samples})
\label{cost_opdm}
\end{equation}
where one needs $(L/2)^4$ loops to estimate the determinant (\ref{slater_opdm})
to find each element of the unitary matrix $\langle [j]|[\alpha]\rangle$; there are $N$ of such elements,
then multiplying with this
the coefficients $\{a_{[j]}\}$ in the computational basis to find the coefficients $\{a_{[\alpha]}\}$'s.
For $L=16$ and $N\simeq 10^4$
the total complexity (\ref{cost_opdm}) is roughly
on the same order of 
the one for diagonalizing the Hamiltonian (\ref{H_mn}), which is $\sim N^3$. 

\begin{figure}
\includegraphics[width=70mm]{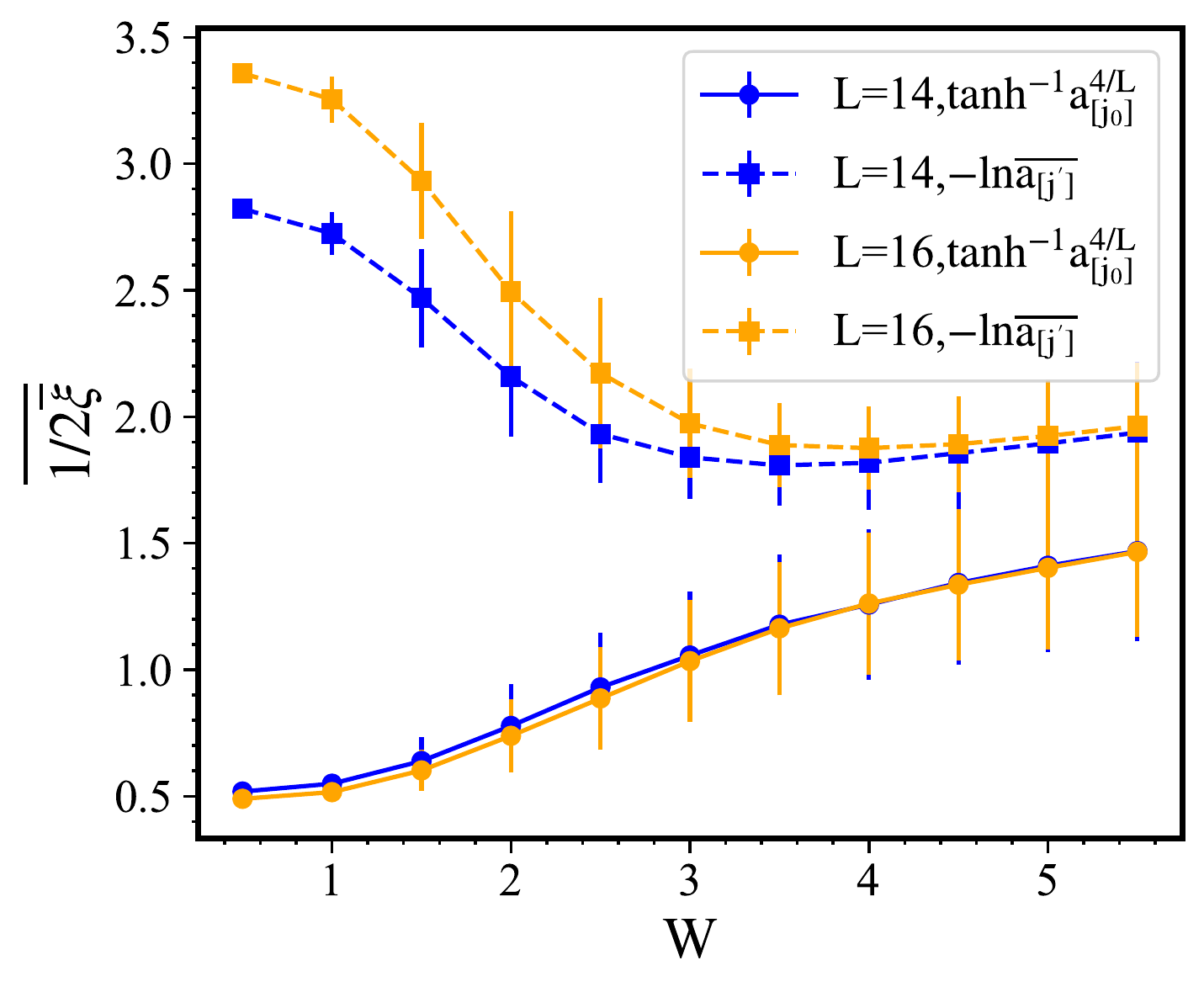}
\caption{
Check of Eqs. (\ref{a_j0}), (\ref{a_j1}) and 
of the underlying hypothesis.
The inverse of the typical localization length,
$1/(2\overline{\xi})$ estimated through
Eqs. (\ref{a_j0b}) and (\ref{a_j1b})
are plotted as a function of $W$
and compared.
}
\label{check}
\end{figure}

\subsection{Anderson localization orbital basis}

In Sec. II-E,
in parallel with Eq. (\ref{IPR_opdm}),
we have also calculated the Fock-space IPR in AL orbital basis:
\begin{equation}
{\cal IPR}_{[\alpha]}^{(AL)}=
\sum_{[\alpha]} \left| a_{[\alpha]}^{(AL)} \right|^4,
\label{IPR_AL}
\end{equation}
where the coefficients $a_{[\alpha]}^{(AL)}$'s are given in Eq. (\ref{psi_AL}).
The coefficients $a_{[\alpha]}^{(AL)}$'s 
are related to the ones in the computational basis, i.e., to $a_{[j]}$'s
through the relation:
\begin{equation}
a_{[\alpha]}^{(AL)}=\sum_{[j]} 
\ _{AL}\langle[\alpha]|[j]\rangle
a_{[j]},
\label{a_alpha_AL}
\end{equation}
where
$\ _{AL}\langle [\alpha]|[j]\rangle=\langle [j]|[\alpha]\rangle^*_{AL}$
is a complex conjugate of the following Slater determinant:
\begin{eqnarray}
&\langle [j]|[\alpha]\rangle_{AL} =
\nonumber \\
&
\det\left[
\begin{array}{cccc}
\psi_{AL}^{(\alpha_1)*}(j_1) & \psi_{AL}^{(\alpha_2)*}(j_1) & \cdots & \psi_{AL}^{(\alpha_{L/2})*}(j_1) \\
\psi_{AL}^{(\alpha_1)*}(j_2)  & \psi_{AL}^{(\alpha_2)*}(j_2)  & \cdots & \psi_{AL}^{(\alpha_{L/2})*}(j_2) \\
\vdots  & \vdots   &  &\vdots \\
\psi_{AL}^{(\alpha_1)*}(j_{L/2}) & \psi_{AL}^{(\alpha_2)*}(j_{L/2}) & \cdots & \psi_{AL}^{(\alpha_{L/2})*}(j_{L/2}) 
\end{array}
\right].
\label{slater_AL}
\end{eqnarray}
The computational task to find
the coefficients $\{a_{[\alpha]}^{(AL)}\}$
from the ones in the computational basis
through
the unitary transformation [Eqs. (\ref{a_alpha_AL}) and (\ref{slater_AL})]
is on the same order of the ones in the case of OPDM basis.
Here,
the only difference is that 
for a given disorder configuration
one can use the same matrix (\ref{slater_AL}) in the computation of $\{a_{[\alpha]}^{(AL)}\}$
in each sampling of the eigenstate $|\psi\rangle$.
The total complexity of the task in the AL orbital basis is, instead of (\ref{cost_opdm}),
\begin{equation}
N^2\times (L/2)^4 \times 1.
\label{cost_AL}
\end{equation}
Thus, sampling many eigenstates $|\psi\rangle$
is numerically less costly in the AL orbital basis.

\section{Check of Eqs. (\ref{a_j0}), (\ref{a_j1}) and 
of the underlying hypothesis}

Here, we numerically evaluate 
Eqs. (\ref{a_j0}), (\ref{a_j1}), 
and by checking the consistencies of these formulas,
certify
the validity of the underlying assumption that
$|\psi\rangle$
is expressed as a simple product of LIOM orbitals as in Eq. (\ref{beta_0}).
To ease the comparison of Eqs. (\ref{a_j0}) and (\ref{a_j1}), 
let us rewrite Eq. (\ref{a_j0}) as
\begin{equation}
\tanh^{-1} a_{[j_0]}^{4/L}
= \frac{1}{ 2\overline{\xi}}.
\label{a_j0b}
\end{equation}
Similarly,
Eq. (\ref{a_j1}) may be rewritten as
\begin{equation}
-\log a_{[j']} = \frac{1}{ 2\overline{\xi}}.
\label{a_j1b}
\end{equation}
Suppose that
the coefficients $a_{\{n_j\}}$
as given in Eq. (\ref{psi_comp})
found in the computational basis
are ordered in the ascending order of $|a_{\{n_j\}}|$,
then relabeled as $a_\mu$ ($\mu=0,1,2,\cdots$).
It is natural to identify
$a_0$ as $a_{[j_0]}$, and also
$\frac{1}{ L} \sum_{\mu=1}^L a_\mu$ as $a_{[j']}$.
Then, we can explicitly evaluate the left-hand sides of
Eqs. (\ref{a_j0b}) and (\ref{a_j1b}).
In Fig. \ref{check}
these two quantities are plotted as a function of $W$
after ensemble averaging.
One can see that
the two quantities tend to merge
in the regime large $W$: i.e., in the MBL regime,
where
$|\psi\rangle$ is presumed to take the simple product form (\ref{beta_0}).

\section{Phase diagram in the ($W,\epsilon$)-plane}

In six panels of Fig. 3
the behaviors of
multi-fractal dimension $D_2$ and its fluctuation
have been considered
in the vicinity of the center of the energy band:
$\epsilon=0.5$, where
$\epsilon$ has been defined in Eq. (\ref{energy_reg}).
Here, we repeat such analyses
away from the $\epsilon=0.5$ region
and
establish the ``ETH-MBL phase diagram''
in the ($W,\epsilon$)-plane;
see Fig. 8.
In the first two panels [(a) and (b)]
the multi-fractal dimension $D_2$ has been
estimated at different values of $W$ and $\epsilon$
in the 
computational [panel (a)]
and 
in the OPDM [panel (b)]
bases.
A contrasting behavior of $D_2$ in the ETH and MBL regions
shows the location of 
ETH-MBL phase boundary (i.e., the location of mobility edge)
in the ($W,\epsilon$)-plane.
Note that this contrast is much sharper in panel (b), i.e.,
in the OPDM basis
than in panel (a), i.e.,
in the computational basis;
compare the
contrast of dominant colors in the two representative regions.
Thus, one can see that
the use of OPDM basis accentuates the difference of ETH and MBL regions,
leading to a substantial improvement of the 
ETH-MBL phase diagram.
Panels (c) and (d)
show
the standard deviation of $D_2$
in the 
computational [panel (c)]
and 
in the OPDM [panel (d)]
bases.
In the latter 
the standard deviation
of $D_2$ is sharply peaked 
in the ETH-MBL crossover regime.
 

Here, we have sampled $10$ eigenstates close to the target energy region for each disorder realization, 
and have averaged the result over
$10^2$ disorder realizations.
Due to long computational time
the system size has been restricted to $L=14$.

\begin{figure}
\includegraphics[width=85mm]{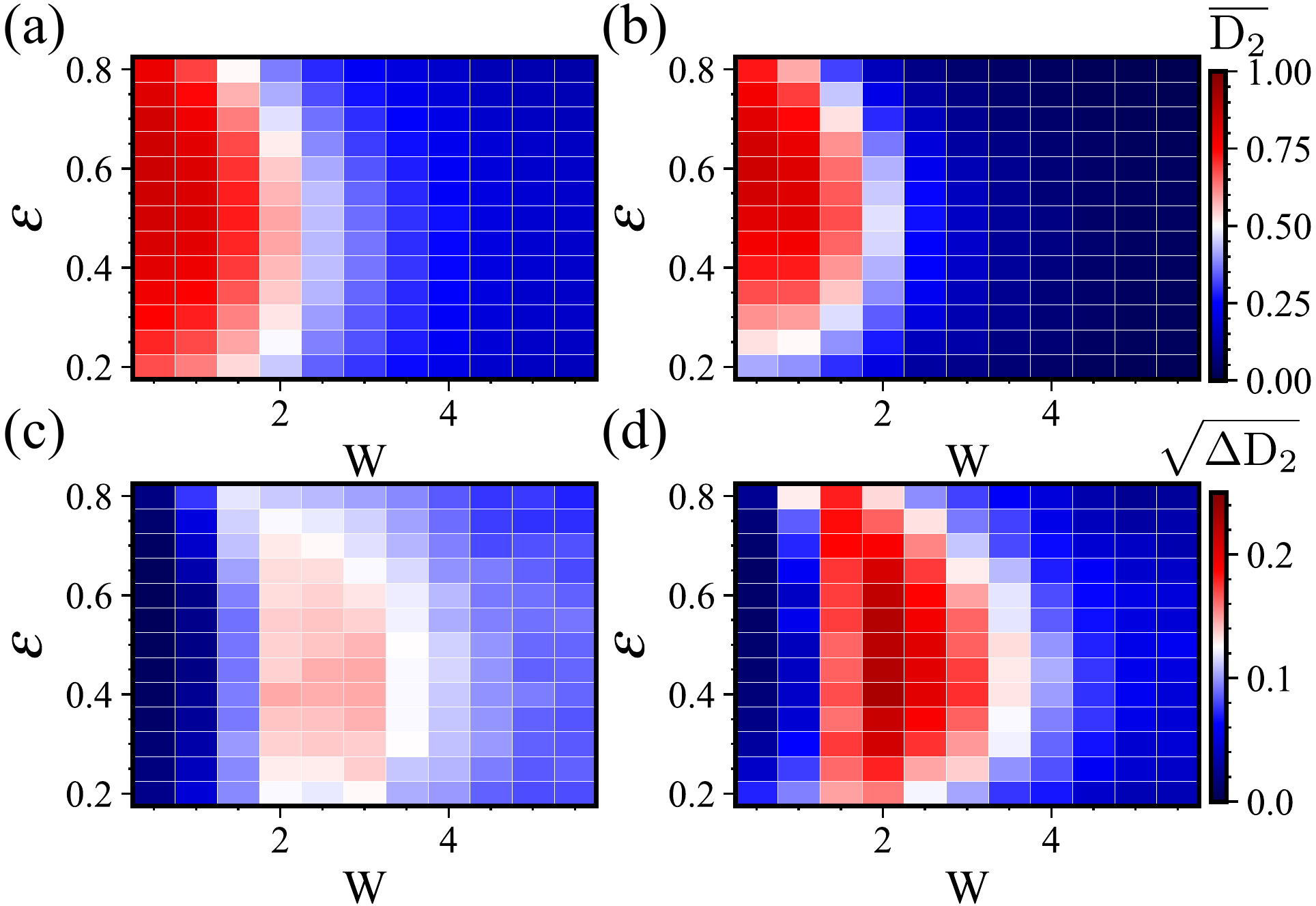}
\caption{
The behavior of
multi-fractal dimension $D_2$ [panels (a),(b)]
and its standard deviation [panels (c),(d)]
away from the mid spectrum ($\epsilon=0.5$);
in the computational [(a),(c)]
and
in the OPDM [(b),(d)]
bases.
}
\label{energy_dep}
\end{figure}

\section{Proof of two formulas in Sec. III}
\subsection{Proof of Eq. (\ref{purity2})}

Let us first recall Eq. (\ref{sigma_j});
\begin{eqnarray}
\langle\hat{\sigma}_\alpha\rangle&=&\langle\psi |\hat{\sigma}_\alpha |\psi\rangle
\nonumber \\
&=&
\left(
\sum_{\mu\in\{\alpha_\uparrow\}} - \sum_{\mu\in\{\alpha_\downarrow\}}
\right)
|a_\mu|^2
\nonumber \\
&=&
\left(
\sum_{\mu_\uparrow} - \sum_{\mu_\downarrow}
\right)
|a_\mu|^2,
\label{sigma_j2}
\end{eqnarray}
where
the notations $\mu_\uparrow$ and $\mu_\downarrow$
have been introduced in Sec. III A
[slightly before Eq. (\ref{sigma_j})].
In the last line we have introduced short-hand notations:
\begin{equation}
\sum_{\mu_\uparrow}=\sum_{\mu\in\{\alpha_\uparrow\}},\ \ \
\sum_{\mu_\downarrow}=\sum_{\mu\in\{\alpha_\downarrow\}}.
\label{mu_updown}
\end{equation}
Using the above notation 
one can reexpress the $j$-th component in the summation
on the r.h.s. of Eq. (\ref{purity}), which defines the local purity,
as
\begin{eqnarray}
\langle\hat{\sigma}_\alpha\rangle^2
&=&
\left\{
\left(
\sum_{\mu_\uparrow} - \sum_{\mu_\downarrow}
\right)
|a_\mu|^2
\right\}^2
\nonumber \\
&=&
\left(
\sum_{\mu_\uparrow, \nu_\uparrow} + \sum_{\mu_\downarrow, \nu_\downarrow}
- \sum_{\mu_\uparrow, \nu_\downarrow} - \sum_{\mu_\downarrow, \nu_\uparrow}
\right)
|a_\mu|^2
|a_\nu|^2,
\label{minus}
\end{eqnarray}
where
$\sum_{\mu_\uparrow, \nu_\downarrow}$ is a short-hand notation:
\begin{equation}
\sum_{\mu_\uparrow, \nu_\downarrow}
=\sum_{\mu\in\{\alpha_\uparrow\}}\sum_{\nu\in\{\alpha_\downarrow\}}.
\end{equation}
Since the many-body state $|\psi\rangle$ is normalized as
\begin{equation}
\sum_\mu |a_\mu|^2=1,
\label{norm}
\end{equation}
the following identity holds:
\begin{eqnarray}
1&=&
\sum_\mu |a_\mu|^2
\sum_\nu |a_\nu|^2
\nonumber \\
&=&
\left(
\sum_{\mu_\uparrow, \nu_\uparrow} + \sum_{\mu_\downarrow, \nu_\downarrow}
+ \sum_{\mu_\uparrow, \nu_\downarrow} + \sum_{\mu_\downarrow, \nu_\uparrow}
\right)
|a_\mu|^2
|a_\nu|^2.
\label{plus}
\end{eqnarray}
Comparing Eqs. (\ref{minus}) and (\ref{plus}),
one finds
\begin{eqnarray}
\langle\hat{\sigma}_\alpha\rangle^2
&=&
1-2\left(
\sum_{\mu_\uparrow, \nu_\downarrow} + \sum_{\mu_\downarrow, \nu_\uparrow}
\right)
|a_\mu|^2
|a_\nu|^2
\nonumber \\
&=&
1-4\sum_{\mu_\uparrow, \alpha_\downarrow} 
|a_\mu|^2
|a_\nu|^2
\nonumber \\
&=&
1-
4\sum_{\mu\in\{\alpha_\uparrow\}}\sum_{\nu\in\{\alpha_\downarrow\}}
|a_\mu|^2
|a_\nu|^2.
\end{eqnarray}
Finally, we plug this expression back into the summation
in the formula (\ref{purity}),
and find Eq. (\ref{purity2}).
In the last step we also note Eq. (\ref{sum_ham}).

\subsection{Proof of  Eq. (\ref{IPR_F2})}

Recalling Eq. (\ref{norm}), and noting the identity:
\begin{equation}
\sum_{\mu,\nu}=\sum_{\mu=\nu}+2\sum_{\mu<\nu},
\end{equation}
one finds
\begin{eqnarray}
1&=&
\sum_{\mu,\nu}
|a_\mu|^2
|a_\nu|^2
\nonumber \\
&=&
\left(
\sum_{\mu=\nu}+2\sum_{\mu<\nu}
\right)
|a_\mu|^2
|a_\nu|^2
\nonumber \\
&=&
\sum_\mu
|a_\mu|^4
-2\sum_{\mu<\nu} 
|a_\mu|^2
|a_\nu|^2.
\end{eqnarray}
Using this,
one can rewrite ${\cal IPR}$ in a form similar to Eq. (\ref{purity2});
i.e., as in Eq. (\ref{IPR_F2}).

\bibliography{ref_MBL6}

\end{document}